\documentclass[11pt]{amsart}
\usepackage{amsfonts,amssymb}
\def\SB{S_B}
\def\SC{S_C}

\def\cosing{C} %C(1)
\def\bbeta{\boldsymbol{\beta}}
\def\ggamma{\boldsymbol{\gamma}}
\def\pphi{\boldsymbol{\varphi}}
\def\Ker{\mathop{\rm Ker}\nolimits}
\def\Im{\mathop{\rm Im}\nolimits}

\newcommand{\UParrow}[1]{\bigm\uparrow{\scriptstyle\!\!#1}\kern-12pt}

\def\ellch{{\mathsf{l}}_{\mathrm{ch}}}

\def\ketAA#1{\mathchoice{{\left|{#1}\right\rangle}_{\bar AA}}%
  {|{#1}\rangle_{\bar AA}}{|{#1}\rangle_{\bar AA}}{|{#1}\rangle_{\bar
      AA}}}

\def\ketbc#1{\mathchoice{{\left|{#1}\right\rangle}_{bc}}%
  {|{#1}\rangle_{bc}}{|{#1}\rangle_{bc}}{|{#1}\rangle_{bc}}}

\def\ketBG#1{\mathchoice{{\left|{#1}\right\rangle}_{\beta\gamma}}%
  {|{#1}\rangle_{\beta\gamma}}{|{#1}\rangle_{\beta\gamma}}{|{#1}\rangle_{\beta\gamma}}}

\def\ketBC#1{\mathchoice{{\left|{#1}\right\rangle}_{BC}}%
  {|{#1}\rangle_{BC}}{|{#1}\rangle_{BC}}{|{#1}\rangle_{BC}}}
\def\ketbc#1{\mathchoice{{\left|{#1}\right\rangle}_{bc}}%
  {|{#1}\rangle_{bc}}{|{#1}\rangle_{bc}}{|{#1}\rangle_{bc}}}

\makeatletter
 \@addtoreset{equation}{section}
\makeatother

\def\double#1{#1,#1}

\newcommand{\cH}{\mathcal{H}}

\newcommand{\N}[1]{N\!=\!#1}
\newcommand{\SL}[1]{s\ell(#1)}
\newcommand{\tSL}[1]{\widehat{s\ell}(#1)}

\def\NPB{Nucl.\ Phys.\ B}
\def\PLB{Phys.\ Lett.\ B}

\def\CMP{Commun.\ Math.\ Phys.}
\def\IJMPA{Int.\ J.\ Mod.\ Phys.\ A}

\newcommand{\oC}{\mathbb{C}}
\newcommand{\oG}{\mathbb{G}}
\newcommand{\oN}{\mathbb{N}}
\newcommand{\oQ}{\mathbb{Q}}

\newcommand{\oZ}{\mathbb{Z}}

\newtheorem{Thm}{Theorem}[section]
\newtheorem{Lemma}[Thm]{Lemma}

\newtheorem*{cor}{Corollary}

\theoremstyle{definition}
\newtheorem{Dfn}[Thm]{Definition}
\newtheorem{Rem}[Thm]{Remark}%[section]

\newenvironment{prf}{%
  \noindent{\sc Proof.}}%
{\noindent{\nolinebreak\hfill\mbox{\rule{.5em}{.5em}}\,}\par\medskip}

\def\cE{\mathcal{E}}

\def\cG{\mathcal{G}}
\def\cH{\mathcal{H}}

\def\cL{\mathcal{L}}

\def\cQ{\mathcal{Q}}

\def\cT{\mathcal{T}}
\def\cU{\mathcal{U}}

\def\smm{\mathfrak}
\def\smA{{\smm{A}}}
\def\smB{{\smm{B}}}
\def\mG{{\smm{G}}}
\def\mV{{\smm{V}}}  %<----- topological module
\def\mU{{\smm{U}}}  %<----- massive module             WHY ALL FRAKTUR?
\def\mK{{\smm{K}}}  %<----- a unitary \N2 module

\def\smV{{\smm{V}}}
\def\pmV{\widehat{\smm{V}}}  % perverted modules
\def\pmU{\widehat{\smm{U}}}  % differently perverted module
\def\dmV{\widetilde{\smm{V}}}
\def\dmU{\widetilde{\smm{U}}}

\def\mH{\smm{H}}

\newcommand{\ket}[1]{\mathchoice{%
    {\left|{#1}\right\rangle}}{|{#1}\rangle}{|{#1}\rangle}{|{#1}\rangle}}

\newcommand{\kettop}[1]{\mathchoice{%
    {\left|{#1}\right\rangle_{\mathrm{top}}}}{\bigl|{#1}\bigr\rangle_{\mathrm{top}}}{|{#1}\rangle_{\mathrm{top}}}{|{#1}\rangle_{\mathrm{top}}}}

\def\ctop{\mathsf{c}}
\def\Ctop{\mathsf{C}}

\def\hplus{\mathsf{h}^+}
\def\hminus{\mathsf{h}^-}

\def\bar{\overline}

\newcommand{\spfn}[1]{\mathop{\mathsf{\bar U}_{#1}}}

\def\hw{highest-weight}

\def\tensor{\otimes}
\renewcommand{\d}{\partial}
\renewcommand{\atop}[2]{\genfrac{}{}{0pt}{}{#1}{#2}}

\begin{document}
\hfuzz=1pt \addtolength{\baselineskip}{4pt}

\title[Free-Field $\N2$ Resolutions ]{\hfill{
    \lowercase{\tt hep-th/9810059}}\\[12pt]
  Free-Field Resolutions of the Unitary $\N2$ Super-Virasoro
  Representations}

\author{B.~L.~Feigin}
\address{Landau Institute for Theoretical Physics, Russian
  Academy of Sciences}
\author{A.~M.~Semikhatov}
\address{Tamm Theory Division, Lebedev Physics Institute, Russian
  Academy of Sciences}

\begin{abstract}
  We construct free-field resolutions of unitary representations of
  the $\N2$ superconformal algebra.  The irreducible representations
  are singled out from free-field spaces as the cohomology of
  fermionic screening operators.  We construct and evaluate the
  cohomology of the resolution associated with one fermionic screening
  (which is related to the representation theory picture of
  ``gravitational descendants''), and a {\it butterfly\/} resolution
  associated with two fermionic screenings.
\end{abstract}

\maketitle

\setcounter{tocdepth}{3}
\vspace*{-12pt}
\begin{center}
  \parbox{.7\textwidth}{
    {\footnotesize
      \tableofcontents}
    }
\end{center}
\vspace*{-24pt}

\thispagestyle{empty}

%\vspace*{-6pt}
\addtolength{\parskip}{2pt}

\section{Introduction}
As is well known, free-field constructions (``bosonisations'') of
infinite-dimensional algebras alter the embedding structure of
Verma-like modules~\cite{[FF]}; resolutions of irreducible
representations are also changed, the classical example being that of
the BGG resolution~\cite{[BGG]} of irreducible Virasoro
representations replaced by the Felder resolution~\cite{[F]}.  The
resolutions associated with bosonisations are related to the screening
operators existing in free-field representation spaces (in fact, to
the corresponding quantum-group structure)~\cite{[F],[BF],[FFr],[BMP]}.

Several free-field constructions of the $\N2$ superconformal extension
of the Virasoro algebra are known. One of these~\cite{[Fre++],[W-LG]}
has been used in the analysis of the Landau--Ginzburg (LG)
models~\cite{[W-LG],[KYY],[NW]}.  There also exist a
bosonisation~\cite{[MSS],[OS],[Ito]} with two fermionic screening
operators, and the $\N2$ construction~\cite{[GS2],[BLNW]} realised in
the bosonic string (although the latter is not necessarily a {\it
  free\/}-field realisation).  Despite the wide use of free-field
constructions, however, the corresponding resolutions of irreducible
$\N2$ representations have not been written out (resolutions of
irreducible $\N2$ representations in terms of {\it Verma\/} modules
were constructed in~\cite{[FSST]}).  In this paper, we fill this gap
for unitary $\N2$ representations~\cite{[BFK]} by constructing
\hbox{resolutions associated with {\it fermionic\/} screenings}.

The first resolution that we construct is in the space of a bosonic
($bc$) and a fermionic ($\beta\gamma$) ghost systems (which is a
particular case of the realisation used in the LG
context~\cite{[W-LG],[KYY],[NW]}, the $\N2$ supersymmetry in the
$bc\beta\gamma$ system being known since~\cite{[FMS]}).  The
resolution is of a ``linear'' two-sided structure, and in this respect
is similar to the $\tSL2$ resolution~\cite{[BF],[FFr]} associated with
the Wakimoto bosonisation.  However, it is {\it not\/} the image under
the $\tSL2\leftrightarrow\N2$ correspondence~\cite{[FST],[DLP]} of the
known $\tSL2$ resolution; instead, its $\tSL2$ counterpart is
constructed of {\it twisted\/} (spectral-flow transformed)
modules.\footnote{We systematically refer to the modules transformed
  by the spectral flow as {\it twisted\/} modules,
  see~\cite{[FST],[ST4],[FSST]}.}  A half of the $\N2$ resolution
consists of the modules generated from the gravitational
descendants~\cite{[Losev],[EKYY],[LW]} of the \hw{} vector in the
cohomology.

Another free-field realisation~\cite{[MSS],[OS],[Ito]} that we
consider here involves two screening operators, which are both
fermions.  These give rise to the {\it butterfly resolution\/}
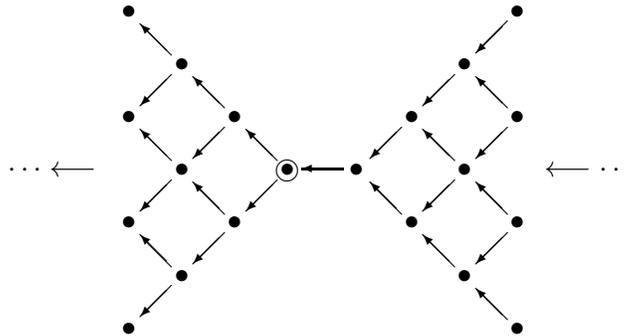
\begin{equation}\label{butterfly-0}
  \unitlength=1pt
  \begin{picture}(400,130)
    \put(20,-20){
      \put(70,80){$\cdots\longleftarrow{}$}
      \put(270,80){${}\longleftarrow\cdots$}
      \multiput(200,80)(20,20){3}{
        \put(20,20){$\bullet$}
        \put(19,19){\vector(-1,-1){12}}
        }
      \multiput(220,60)(20,20){2}{
        \put(20,20){$\bullet$}
        \put(19,19){\vector(-1,-1){12}}
        \put(19,26){\vector(-1,1){12}}
        }
      \multiput(240,40)(20,20){1}{
        \put(20,20){$\bullet$}
        \put(19,19){\vector(-1,-1){12}}
        \put(19,26){\vector(-1,1){12}}
        }
      \multiput(200,80)(20,-20){3}{
        \put(20,-20){$\bullet$}
        \put(19,-14){\vector(-1,1){12}}
        }
    %center
      %\put(201.7,82.5){\circle{8}}
      \put(175.7,82.5){\circle{8}}
      \put(173,80){
        \put(0,0){$\bullet$}
        \put(24,3){\vector(-1,0){16}}
        \put(26,0){$\bullet$}
        }
      \multiput(173,80)(-20,20){3}{
        \put(-20,20){$\bullet$}
        \put(-1,6){\vector(-1,1){12}}
        }
      \multiput(153,60)(-20,20){2}{
        \put(-20,20){$\bullet$}
        \put(-1,6){\vector(-1,1){12}}
        \put(-1,39){\vector(-1,-1){12}}
        }
      \multiput(133,40)(-20,20){1}{
        \put(-20,20){$\bullet$}
        \put(-1,6){\vector(-1,1){12}}
        \put(-1,39){\vector(-1,-1){12}}
        }
      \multiput(153,60)(-20,-20){3}{
        \put(19,19){\vector(-1,-1){12}}
        \put(0,0){$\bullet$}
        }
      }
  \end{picture}
\end{equation}
whose two-winged shape seems to be an entirely new structure.  
%The obvious moral is that the structure of the resolution is quite
%sensitive to the ``bosonisation'' used.
In both cases, however, the cohomology is given by the unitary $\N2$
representations (Theorems~\ref{thm:Q0} and~\ref{thm:butterfly}).
Recall that the unitary $\N2$ representations are characterised by the
relation
\begin{equation}\label{G(z)-relation}
  \d^{p-2}\cG(z)\dots\d\cG(z)\,\cG(z)=0\,,
\end{equation}
where $\cG(z)$ is one of the two fermionic fields entering the $\N2$
superconformal algebra.  This gives an alternative way to show that
the cohomology is a sum of unitary representations: it suffices to
check the action of the $\N2$ algebra and to verify
that~\eqref{G(z)-relation} is satisfied (in the $bc\beta\gamma$
realisation, for example, the latter condition holds because the
$\gamma(z)$ field satisfies $\gamma^{p-1}=0$ in the cohomology).

The general features that are important in the analysis of $\N2$
representations (and which are absent in the Virasoro and $\N1$
superconformal algebras) are the spectral flow transform~\cite{[SS]}
and the appearance of {\it
  twisted\/}---spectral-flow-transformed---(sub)modules, even if one
starts with an ``untwisted'' module~\cite{[ST4]}.  The spectral flow
maps between different unitary $\N2$ representations, the length of
the orbit being~$p$ (or~$p/2$ in a certain special case of
even~$p$)~\cite{[FST],[FSST]}, where $p=(2),3,4,\dots$ parametrises
the central charge.  Thus, all of the unitary representations can be
obtained by applying the spectral flow to only $[p/2]$ representatives
of the spectral flow orbits, and similarly for the resolutions.

As with all the ``invariant'' structures pertaining to $\N2$
representations, the resolutions we construct have their $\tSL2$
counterparts.  We will comment on how the two resolutions are mapped
into the Wakimoto bosonisation~\cite{[Wak],[FFr-0],[GMMOS]} of
$\tSL2$.  {}From a more general perspective, different free-field
realisations are particular cases of the three-boson realisation of
the $\N2$ (or $\tSL2$) algebra.  We do not attempt here to analyse the
generic three-boson realisation resolutions, which should include
(i)~the case of a bosonic and a fermionic screening that make up the
nilpotent subalgebra of $s\ell(2|1)_q$, and another bosonic screening
commuting with the first two; as regards the ``$s\ell(2|1)_q$-pair,''
we hope to consider it in the future, while taking a bosonic and a
fermionic screening that commute with each other leads eventually to
the Felder-type resolution, since the fermionic screening singles out
a $\beta\gamma$ system, after which the bosonic screening acts as
in~\cite{[BF],[FFr]}; \ (ii)~the case of two fermionic screenings and
a non-vertex-operator bosonic screening.  This also has not been
worked out in general, but what we consider in
Sec.~\ref{sec:symmetric} is an important particular case---that of
integrable representations---where the fermionic screenings commute
(and, thus, make it \hbox{possible to consider the resolution with one
  fermionic screening as in Sec.~\ref{sec:ghost})}.

\smallskip

In Sec.~\ref{sec:general}, we recall some basic facts about the $\N2$
superconformal algebra and its unitary representations. In
Sec~\ref{sec:ghost}, we consider the $bc\beta\gamma$ realisation of
the $\N2$ algebra.  We evaluate the BRST cohomology of the fermionic
screening and show how the corresponding resolution gives rise to
gravitational descendant fields.  In Sec.~\ref{sec:symmetric}, we
consider the bosonisation in terms of a complex scalar and a fermionic
ghost system and construct the butterfly resolution, whose cohomology
is also given by the unitary $\N2$ representations.  We also discuss
the relation between the two resolutions.
%and comment on their mapping to the $\tSL2$ resolutions
%(Secs.~\ref{sec:jamming} and~\ref{subsec:to-Wak}).

\section{Generalities\label{sec:general}}
\subsection{The $\N2$ algebra and the spectral flow} The $\N2$
superconformal algebra is taken in the basis where the nonvanishing
commutation relations read as
\begin{alignat}{2}\label{topalgebra}
  {[}\cL_m,\cL_n]&=(m-n)\cL_{m+n}\,,&[\cH_m,\cH_n]&=
  \tfrac{\Ctop}{3}m\delta_{m+n,0}\,,\notag\\
  {[}\cL_m,\cG_n]&=(m-n)\cG_{m+n}\,,&
  [\cH_m,\cG_n]&=\cG_{m+n}\,, \notag\\
  {[}\cL_m,\cQ_n]&=-n\cQ_{m+n}\,,&
  [\cH_m,\cQ_n]&=-\cQ_{m+n}\,,\displaybreak[0]\\
  {[}\cL_m,\cH_n]&=
  -n\cH_{m+n}+\tfrac{\Ctop}{6}(m^2+m)\delta_{m+n,0}\,,\kern-80pt
  \notag\\
  \{\cG_m,\cQ_n\}&=
  \cL_{m+n}-n\cH_{m+n}+
  \tfrac{\Ctop}{6}(m^2+m)\delta_{m+n,0}\,,\kern-80pt\notag
\end{alignat}
with $m,~n\in\oZ$.  Here, $\cL_n$ and $\cH_n$ are bosonic, and $\cG_n$
and $\cQ_n$, fermionic elements.  In what follows, we do not
distinguish between the central element $\Ctop$ and its
eigenvalue~$\ctop$, which we assume to be $\ctop\neq3$ and parametrise
as $\ctop=3(1-\frac{2}{t})$ with~$t\in\oC\setminus\{0\}$.  In the
unitary case, we write $t=p=2,3,\dots$ (where $p=2$ leads to the
trivial representation).  To describe the same algebra in terms of
{\it currents\/} and operator products, we define
$\cQ(z)=\sum_{n\in\oZ}\cQ_nz^{-n-1}$,
$\cG(z)=\sum_{n\in\oZ}\cG_nz^{-n-2}$,
$\cT(z)=\sum_{n\in\oZ}\cL_nz^{-n-2}$, and
$\cH(z)=\sum_{n\in\oZ}\cH_nz^{-n-1}$.

In the basis chosen in~\eqref{topalgebra}, the spectral
flow~\cite{[SS]} acts as
\begin{gather}
  \spfn{\theta}:
  \begin{aligned}
    \cL_n&\mapsto\cL_n+\theta\cH_n+\tfrac{\ctop}{6}(\theta^2+\theta)
    \delta_{n,0}\,,&\qquad{}
    \cH_n&\mapsto\cH_n+\tfrac{\ctop}{3}\theta\delta_{n,0}\,,\\
    \cQ_n&\mapsto\cQ_{n-\theta}\,,&{}\cG_n&\mapsto\cG_{n+\theta}\,.
  \end{aligned}
  \label{U}
\end{gather}
For $\theta\in\oZ$, these transformations are automorphisms of the
algebra.  Allowing $\theta$ to be half-integral in~\eqref{U}, we
obtain the isomorphism between the Ramond and Neveu--Schwarz sectors.
We refer to the modules subjected to the action of the spectral flow
as {\it twisted\/} modules; the algebra acts on a twisted module
according to the standard prescription that ``a given generator acts
as the spectral-flow transformed generator acts on the original
module.''

\subsection{Unitary representations of the $\N2$
  algebra\label{sec:unitary}} We now briefly review,
following~\cite{[FSST]}, several facts about the unitary $\N2$
representations.  These are the irreducible quotients of the twisted
topological\footnote{{\it Chiral\/} modules in a different
  nomenclature~\cite{[LVW]}.} Verma modules (see the Appendix
or~\cite{[ST4],[SSi]} for more details)
$\smV_{\hplus(r,1,p),p;\theta}$, where the variable parametrising the
central charge is $t=p\in\oN+1$,\footnote{$\oN=\{1,2,\ldots\}$ and,
  for the future use, $\oN_0=\{0,1,2,\ldots\}$.} \ $\hplus$ is defined
in~\eqref{hplusminus}, $r$ is an integer such that $1\leq r\leq p-1$,
and $\theta$ is the twist (see~Definition~\ref{dfn:top}).  Although
the Verma module can be taken with any integral twist $\theta$, the
unitary representations are periodic with period~$p$ (i.e., acting
with the spectral flow transform with $\theta=p$ gives an isomorphic
representation):
\begin{equation}\label{periodicity}
  \mK_{r,p;\theta+p}\approx\mK_{r,p;\theta}\,.
\end{equation}
For the unitary representations, the twist $\theta$ can therefore be
considered mod~$p$; thus, the unitary representations are labelled by
\begin{equation}\label{all-unitary}
  \mK_{r,p;\theta}\,,\qquad 1\leq r\leq p-1\,,~\theta\in\oZ_p\,.
\end{equation}
Among these, {\it there are only $p(p-1)/2$ non-isomorphic unitary
  representations}, since there exist the $\N2$ isomorphisms
\begin{equation}\label{identification}
  \mK_{r,p;\theta+r}\approx\mK_{p-r,p;\theta}\,,\qquad
  1\leq r\leq p-1\,,\quad\theta\in\oZ_p\,.
\end{equation}
Thus, in order to count each unitary representation once, we can, for
example, take $1\leq r\leq[p/2]$ with the full range of~$\theta$,
$0\leq\theta\leq p-1$, or allow $1\leq r\leq p-1$ with
$0\leq\theta\leq r-1$.  There is a special periodicity property
applying to the representations $\mK_{r,2r;\theta}$, for which the
period is half that of Eq.~\eqref{periodicity}: $\mK_{r,2r;\theta + r}
\approx \mK_{r,2r;\theta}$.

As noted in the Introduction, the structure of all the unitary
representations is described once this is done for a representative of
each spectral flow orbit.  In the next section, for example, we
concentrate on the representations $\mK_{r,p;r-1}$, which in terms of
the Verma modules are the quotients (see~\eqref{standard-emb})
\begin{equation}\label{Verma-quotient}
  \mK_{r,p;r-1}=\mV_{\frac{1-r}{p},p;r-1}\bigm/
    \bigl(\mV_{\frac{r+1}{p}-2,p;p-1}+\mV_{\frac{r+1}{p},p;-1}\bigr)\,.
\end{equation}
The untwisted representations will also be denoted by
$\mK_{r,p}\equiv\mK_{r,p;0}$.

\section{The ghost realisation\label{sec:ghost}} The free-field
realisation of the $\N2$ algebra that we consider here has a known
relation to the $A_{p-1}$ LG models.  Given a sum
$W(\gamma_1,\dots,\gamma_n)$ of the $A$-series LG potentials, one
constructs the Koszul differential associated with the LG ``equations
of motion'' $\d W/\d\gamma_i=0$~\cite{[Fre++],[W-LG],[KYY]}; for an
individual $A_{p-1}$ model, this differential is
\begin{equation}\label{differential}
  \oQ_0=\tfrac{1}{2\pi i}\oint c\,\gamma^{p-1}\,,
\end{equation}
which, according to the standard BRST ideology, imposes the constraint
\begin{equation}\label{gamma-constraint}
  \gamma(z)^{p-1}\approx0\,.
\end{equation}
The $c$ and $\gamma$ fields involved in~\eqref{differential} are
viewed as the respective halves of a $bc$ (fermionic) and a
$\beta\gamma$ (bosonic) first-order systems with the operator
products
\begin{equation}\label{bcbetagamma}
  b(z)\,c(w)=\frac{1}{z-w}\,,\qquad
  \beta(z)\,\gamma(w)=\frac{-1}{z-w}\,.
\end{equation}
Then~\cite{[Fre++],[W-LG]} $\oQ_0$ commutes with (is the screening of)
the $\N2$ algebra~ realised in terms of the {\it currents\/} $\cQ(z)$
etc. as
\begin{equation}\label{bcBG}
  \begin{split}
    \cQ &= -\beta c\,,\\
    \cG &= -\tfrac{1}{p} \gamma\d b -       % divided by 2
    (\tfrac{1}{p} - 1) \d\gamma b\,,\\    %
    \cH &= -\tfrac{1}{p}\beta\gamma + (1 - \tfrac{1}{p})bc\,,\\
    \cT &= \tfrac{1}{p}\d b\,c - (1 - \tfrac{1}{p}) b\,\d c +
    \tfrac{1}{p}\d\beta\,\gamma - (1-\tfrac{1}{p})\beta\,\d\gamma\,.
  \end{split}
\end{equation}
This algebra helps organise the cohomology of~$\oQ_0$, which, as could
be expected~\cite{[W-LG],[KYY]}, is given by a sum of unitary $\N2$
representations (Eq.~\eqref{sum-unitary}).  To this end, we construct
the resolution, which will in turn require analysing the
representations of~\eqref{bcBG} in some detail.  We assume, as before,
$p\in\oN+1$.  We now digress to fix the conventions regarding ghost
systems.

\subsection{Modding, etc., of the ghost systems} We will consider the
$bc$ and $\beta\gamma$ fields with a fractional modding, i.\,e., with
``twisted boundary conditions.''  We assume the point of view that
twisted boundary conditions on fermions give rise to a fractional
fermion number of the vacuum.  For first-order fields of a
(half-)integer spin~$\lambda$, it is convenient to introduce the
Fourier modes as $b(z)=\sum b_nz^{-n-\lambda}$ and $c(z)=\sum
c_nz^{-n-1+\lambda}$, and impose the annihilation conditions on the
$n$th-picture~vacuum~as
\begin{equation}
  b_{m+1-\lambda-n}\,\ketbc{n}=0\,,\quad
  c_{m+\lambda+n}\,\ketbc{n}=0\,,\qquad m=0,1,2,\dots.
\end{equation}
Then it follows that $(bc)_0\,\ketbc{n}=n\ketbc{n}$.  We continue
these relations to the case of rational $\lambda$ and the picture
number~$n$.

In what follows, the $bc$ and $\beta\gamma$ picture numbers $n$ and
$\nu$ will be fractions such that
\begin{equation}\label{pictures:extra}
  p\nu\in\oZ\,,\qquad \nu-n\in\oZ\,.
\end{equation}
The ghost systems in~\eqref{bcBG} have conformal spin
\begin{equation}
  \lambda = 1 - \tfrac{1}{p}\,.
\end{equation}
Thus, the modding of the ghost fields is taken as
\begin{equation}\label{modding}
  \begin{split}
    &b_m,~m\in\tfrac{1}{p}-n+\oZ\,,
    \qquad c_m,~m\in-\tfrac{1}{p}+n+\oZ\,,\\
    &\beta_m,~m\in\tfrac{1}{p}-\nu+\oZ\,,\qquad
    \gamma_m,~m\in-\tfrac{1}{p}+\nu+\oZ\,.
  \end{split}
\end{equation}

We define the $bc$ module $\Lambda_\lambda(n)$ with the cyclic vector
$\ketbc{n}$ and the $\beta\gamma$ module $\Xi_\lambda(\nu)$ with the
cyclic vector $\ketBG{\nu}$ subjected to the annihilation conditions
\begin{alignat}{3}
  b_{\geq1-\lambda-n}\,\ketbc{n}&=0\,,\qquad
  &c_{\geq\lambda+n}\,\ketbc{n}&=0\,,\label{bcannihil}
  \displaybreak[0]\\
  \beta_{\geq1-\lambda-\nu}\,\ketBG{\nu}&=0\,,\
  &\gamma_{\geq\lambda+\nu}\,\ketBG{\nu}&=0\,,\label{BGannihil}
\end{alignat}
where $n$ and $\nu$ are as in~\eqref{pictures:extra} and where we use
the convention that $b_{\geq1-\lambda-n}$ means
$(b_{m+1-\lambda-n})_{m\in\oN_0}$ and $c_{\geq\lambda+n}$ means
$(c_{m+\lambda+n})_{m\in\oN_0}$ (here, $\Lambda_\lambda(n)$ and
$\Lambda_\lambda(n')$ are of course isomorphic whenever $n-n'\in\oZ$).
Then,
\begin{align}
  (bc)_0\,\ketbc{n}&=n\ketbc{n}\,,\label{bceigen}\\
  (\beta\gamma)_0\,\ketBG{\nu}&=-\nu\ketBG{\nu}\,.\label{BGeigen}
\end{align}

It follows, in particular, that the modes of the current
$\oQ(z)=c(z)\,\gamma(z)^{p-1}$ read as
\begin{equation*}
  \oQ_j=\sum_{m_1,\dots,m_{p-1}\in\oZ-\frac{1}{p}}
  c_{j-m_1-\dots-m_{p-1}}\gamma_{m_1}\dots\gamma_{m_{p-1}}
\end{equation*}
with $j\in\oZ$, which is consistent because $-m_1-\dots-m_{p-1}\in
(p-1)\frac{1}{p}-(p-1)\nu+\oZ= -\frac{1}{p} + \nu + \oZ =-\frac{1}{p}
+ n + \oZ$ in view of~\eqref{pictures:extra}.

For the $\N2$ representation generated from the vector
$\ketbc{n}\tensor\ketBG{\nu}\in\Lambda_\lambda(n)
\tensor\Xi_\lambda(\nu)$, the $\N2$ spectral flow transform
$\cU_\vartheta$ is realised by changing the $bc$ and $\beta\gamma$
pictures as
\begin{equation}\label{ghost-spectral}
  \Lambda_\lambda(n)\to\Lambda_\lambda(n - \vartheta\lambda)\,,
  \qquad
  \Xi_\lambda(\nu)\to\Xi_\lambda(\nu + \vartheta(1-\lambda))\,.
\end{equation}
For $\vartheta\notin p\oZ$, pictures are changed by non-integer
numbers, thereby leading to an inequivalent representation even in the
$bc$ sector alone (while the spectral flow with $\vartheta\in p\oZ$,
although changing the $\beta\gamma$ representation, induces an
isomorphism on the cohomology, as we will see).  Anyway, this
realisation of the $\N2$ spectral flow allows us to fix the overall
twist in an arbitrary way until the very end, when the spectral flow
transform with any desired $\vartheta$ can be applied to the
free-field representations.  In what follows, it will be convenient to
fix the twist as $r-1$; we will arrive \hbox{at the unitary
  representation $\mK_{r,p;r-1}$,~Eq.~\eqref{Verma-quotient}}.

\subsection{The structure of the ghost realisation: ``gravitational
  descendants'' and the resolution\label{sec:ghost-structure}} We now
describe a complex of $\N2$ representations on the $bc\beta\gamma$
space $\Lambda_\lambda(n)\tensor\Xi_\lambda(\nu)$ whose cohomology
gives the unitary $\N2$ representations.  Recall that the
$bc\beta\gamma$ system is an essential ingredient of the conformal
field theory description of topological
gravity~\cite{[DVV-notes],[LLS]}, where this system gives rise to the
gravitational descendants~\cite{[Losev],[EKYY],[LW]} of primary
fields.  The representation-theoretic picture of the gravitational
descendants will also be seen from the complex
(Remark~\ref{rem:grav-desc}).

\medskip

\noindent{\it Notation for the modules}. A Verma module
$\mU_{h,\ell,p;\theta}$ (see~\eqref{verma}--\eqref{masshw-2}) is
uniquely characterised by $(h,\ell,p;\theta)$, i.e., by the \hw{}
conditions (including the Cartan eigenvalues) satisfied by the \hw{}
vector.  In the free-field realisation, on the contrary, a given
module is not characterised by the \hw{} conditions satisfied by the
vector(s) from which the module is generated.  However, indicating the
values of $h$, $\ell$, and $\theta$ is still very useful in the
analysis of mappings between modules.  We will thus use the notation
$\pmU_{h,\ell,p;\theta}\left(M,N\right)$ for the module generated from
$\ketbc{M}\tensor\gamma_{\lambda+\nu-1}^{N}\ketBG{\nu}$ for~$N\geq0$,
or from $\ketbc{M}\tensor\beta_{-\lambda-\nu}^{-N}\ketBG{\nu}$ for
$N<0$, once this vector satisfies the same \hw{} conditions as the
twisted massive \hw{} vector $\ket{h,\ell,p;\theta}$; we also write
$\pmV_{h,p;\theta}\left(M,N\right)$ for the module generated from
$\ketbc{M}\tensor\gamma_{\lambda+\nu-1}^{N}\ketBG{\nu}$, $N\geq0$ (or
$\ketbc{M}\tensor\beta_{-\lambda-\nu}^{-N}\ketBG{\nu}$, $N<0$) when
this vector satisfies the same \hw{} conditions as the twisted
topological \hw{} vector $\kettop{h,p;\theta}$.  Moreover, we will in
some cases omit the arguments $(M,N)$ altogether, simply indicating
the ghost realisation of the \hw{} vector of the module the first time
the module appears.
\begin{Thm}\label{thm:Q0}
  Let $\mG_{n,\nu,p}=\Lambda_{\lambda}(n)\tensor \Xi_{\lambda}(\nu)$,
  $\lambda=1-\frac{1}{p}$, be the ghost representation space defined
  in accordance with~\eqref{bcannihil}--\eqref{BGeigen}, where
  $n,\,\nu\in\frac{1}{p}\,\oZ$, \ $\nu-n\in\oZ$, and
  \begin{equation*}
    1 + (\nu+1)(1-p)\leq n\leq\nu(1-p)\,.
  \end{equation*}
  Then there is a complex of $\N2$ representations on $\mG_{n,\nu,p}$
  \begin{multline}\label{complex-nu}
    \ldots\xrightarrow{\oQ_0}
    \pmU_{\frac{r+1}{p}-m-1,0,p;p-1+\nu p}
    \xrightarrow{\oQ_0}\dots\xrightarrow{\oQ_0}
    \pmU_{\frac{r+1}{p}-2,0,p;p-1+\nu p}
    \xrightarrow{\oQ_0}
    \pmU_{\frac{r+1}{p},0,p;\nu p-1}
    \xrightarrow{\oQ_0}\\
    \xrightarrow{\oQ_0}\pmU_{\frac{r+1}{p}+1,0,p;\nu
      p-1}\xrightarrow{\oQ_0}\ldots
    \xrightarrow{\oQ_0}
    \pmU_{\frac{r+1}{p}+m,0,p;\nu p-1}
    \xrightarrow{\oQ_0}\ldots
  \end{multline}
  where $r=1 + \nu(1-p) - n$, the modules
  $\pmU_{\frac{r+1}{p}+m,0,p;\nu p-1}$ with $m\geq0$ are generated by
  \hbox{the $\N2$} generators~\eqref{bcBG} from the states\; $\ketbc{1
    + \nu(1-p)}\tensor\gamma_{\lambda+\nu-1}^{mp+r} \ketBG{\nu}\in
  \Lambda_{\lambda}(n)\tensor\Xi_{\lambda}(\nu)$,\; and the modules
  $\pmU_{\frac{r+1}{p}-m,0,p;p-1+\nu p}$, $m\geq2$, from $\ketbc{2-p +
    \nu(1-p)}\tensor\beta_{-\lambda-\nu}^{mp-r-1}\ketBG{\nu}$.  The
  cohomology of~\eqref{complex-nu} is concentrated at the term
  $\pmU_{\frac{r+1}{p},0,p;\nu p-1}(1 + \nu(1-p),r)$ and is given by
  the unitary $\N2$ representation~$\mK_{r,p;\theta}$ with
  \begin{equation*}
    r=1 + \nu(1-p) - n\,,\quad \theta=\nu - n\,.
  \end{equation*}
\end{Thm}
Since $\Lambda_\lambda(n)\approx\Lambda_\lambda(n')$ whenever
$n-n'\in\oZ$, we obtain
\begin{cor}
  The cohomology of $\oQ_0$ on $\mG_{n,\nu,p}$ is given by the direct
  sum of unitary $\N2$ representations
  \begin{equation}\label{sum-unitary}
    \bigoplus_{r=1}^{p-1}\mK_{r,p;r-1+\nu p}\,.
  \end{equation}
\end{cor}

The data in the conditions of the Theorem are invariant under the
shifts
\begin{equation}
  n\mapsto{}n+ a(1-p)\,,\qquad
  \nu\mapsto{}\nu + a\,,
\end{equation}
which change the twist as $\theta\mapsto\theta+pa$. For $a\in\oZ$,
this induces the isomorphism~\eqref{periodicity} of the unitary
representations.  For a {\it non-integral\/}~$a\in\frac{1}{p}\oZ$,
such a shift leads to another unitary $\N2$ representation, however
the difference amounts to the overall spectral flow transform, which
is applied to the ghost spaces in accordance
with~\eqref{ghost-spectral}.  Thus, the theorem is {\it equivalent\/}
to its $\nu=0$-case, \hbox{which reads as follows}.
\addtocounter{Thm}{-1}
\def\theThm{\thesection.\arabic{Thm}${}^{\nu=0}$}
\begin{Thm} Let $1\leq r\leq p-1$ and let
  $\mG_{1-r,0,p}=\Lambda_{\lambda}(1-r)\tensor \Xi_{\lambda}(0)\approx
  \Lambda_{\lambda}(0)\tensor \Xi_{\lambda}(0)$,
  $\lambda=1-\frac{1}{p}$, be the ghost representation space.  Then
  there is a complex
  \begin{multline}\label{complex}
    \ldots\rightarrow
    \pmU_{\frac{r+1}{p}-m,0,p;p-1}
    \rightarrow\dots\rightarrow
    \pmU_{\frac{r+1}{p}-2,0,p;p-1}\rightarrow\\
    \rightarrow
    \pmU_{\frac{r+1}{p},0,p;-1}
    \rightarrow\pmU_{\frac{r+1}{p}+1,0,p;-1}\rightarrow
    \ldots\rightarrow
    \pmU_{\frac{r+1}{p}+m,0,p;-1}
    \rightarrow\ldots
  \end{multline}
  where the modules $\pmU_{\frac{r+1}{p}+m,0,p;-1}$, $m\geq0$, are
  generated by $\N2$ generators~\eqref{bcBG} from the states
  $\ketbc{1}\tensor\gamma_{\lambda-1}^{mp+r}\ketBG{0}$, and the
  modules $\pmU_{\frac{r+1}{p}-m,0,p;p-1}$, $m\geq2$, from
  $\ketbc{2-p}\tensor\beta_{-\lambda}^{mp-r-1}\ketBG{0}$.  The
  cohomology of~\eqref{complex} is concentrated at the term
  $\pmU_{\frac{r+1}{p},0,p;-1}(1,r)$ and is given by the unitary
  representation~$\mK_{r,p;r-1}$.

  An equivalent form of~\eqref{complex} is
  \begin{multline}\label{complex-top}
    \ldots\rightarrow
    \pmV_{\frac{1-r}{p}+m-1,p;r-(m-1)p-1}((m-1)p-r+1,0)
    \rightarrow\dots\rightarrow
    \pmV_{\frac{1-r}{p}+1,p;r-p-1}(p-r+1,0)\rightarrow\\
    \rightarrow
    \pmU_{\frac{r+1}{p},0,p;-1}
    \rightarrow\pmU_{\frac{r+1}{p}+1,0,p;-1}\rightarrow
    \ldots\rightarrow
    \pmU_{\frac{r+1}{p}+m,0,p;-1}
    \rightarrow\ldots
  \end{multline}
\end{Thm}
\def\theThm{\thesection.\arabic{Thm}}
\begin{Rem}\label{rem:grav-desc} In~\eqref{complex}, the modules
  $\pmU_{\frac{r+1}{p}+m,0,p;-1}$, $m\geq0$, are generated from the
  ``{\it gravitational descendant\/}'' states
  \begin{equation*}
    \ketbc{1}\tensor\gamma_{\lambda-1}^{mp+r}\ketBG{0}
    \equiv
    \sigma_{(p)}^m\bigl(
    \ketbc{1}\tensor\gamma_{\lambda-1}^{r}\ketBG{0}\bigr)\,,
    \qquad \sigma_{(p)}=\gamma_{\lambda-1}^p\,,
  \end{equation*}
  which are all $\oQ_0$-trivial except the original one with $m=0$. In
  the topological gravity, the $bc\beta\gamma$ states are tensored
  with primaries from other sectors, which gives the gravitational
  descendants.  As we will see in what follows, the ``tic-tac-toe''
  equations relating the gravitational descendants read as
  \begin{equation}\label{tic-tac-toe}
    \cQ_1\,\ketbc{1}\tensor\gamma_{\lambda-1}^{(m+1)p+r}\,\ketBG{0}=
    (mp+r)\,
    \oQ_0\,\ketbc{1}\tensor\gamma_{\lambda-1}^{mp+r}\,\ketBG{0}
  \end{equation}
  (these are in fact an essential ingredient in the construction of
  the resolution).  Further,
  $\cQ_1\,\ketbc{1}\tensor\gamma_{\lambda-1}^{mp+r}\,\ketBG{0}=
  \ketbc{0}\tensor\gamma_{\lambda-1}^{mp+r-1}\,\ketBG{0}$ is a
  singular vector that satisfies twisted {\it topological\/} \hw{}
  conditions.  As in the LG setting, the $m=0$ state is singled out
  from these because $\gamma_{\lambda-1}$ (which in the invariant
  terms is the top mode of $\gamma$ that does not annihilate the
  vacuum) is raised to the power $0\leq r-1\leq k=p-2$.  The effects
  due to the gravitational dressing have been discussed in different
  languages (see~\cite{[L]} and references therein, in
  particular,~\cite{[BLNW],[DVV],[Losev],[EKYY],[LW]}).
\end{Rem}
To prove the Theorem, we need to analyse the structure of the ghost
realisation.

\smallskip

Consider the $\N2$ module spanned by generators~\eqref{bcBG} acting on
the \hw{} vector $\ketbc{n}\tensor\ketBG{\nu}$. This state satisfies
the same annihilation and eigenvalue equations as the twisted
topological \hw{} state $\kettop{(1-\tfrac{1}{p})\nu + \tfrac{n}{p},
  p; \nu-n}$ (see Definition~\ref{dfn:top}), which we express by
writing
\begin{equation}\label{state}
  \ketbc{n}\tensor\ketBG{\nu}\doteq\kettop{
    (1-\tfrac{1}{p})\nu + \tfrac{n}{p}, p;
    \nu-n}\,.
\end{equation}
If, in accordance with the above, the $\beta\gamma$ system is taken in
the zero picture (and hence the $bc$ picture is integral), we consider
the \hw{} vector
\begin{equation}\label{hw-ghost}
  \ketbc{1-r}\tensor\ketBG{0}\doteq
  \kettop{\tfrac{1-r}{p}, p;r-1}
\end{equation}
and denote by $\pmV_{\frac{1-r}{p}, p;r-1}(1-r,0)\subset
\Lambda_\lambda(1-r)\tensor\Xi_\lambda(0)$, $\lambda=1-\frac{1}{p}$,
the $\N2$ representation generated from it.

The following Lemma is an immediate result of direct calculations; we
give it as a separate statement because it is often used in what
follows.
\begin{Lemma}The state $\ketbc{m}\tensor\gamma_{\lambda-1}^a\ketBG{0}$
  satisfies the same \hw{} conditions as the twisted massive \hw{}
  vector $\ket{\frac{m+a}{p},\frac{a(1-m)}{p},p;-m}$ if $m\neq0$, and
  twisted topological \hw{} conditions~\eqref{annihiltop} for
  $m=0${\rm:}
  \begin{equation}
    \ketbc{m}\tensor\gamma_{\lambda-1}^a\ketBG{0}\doteq
    \left\{
      \begin{aligned}
        {}&\ket{\tfrac{m+a}{p},\tfrac{a(1-m)}{p},p;-m}\,,\quad &
        m\in\oZ\setminus\{0\}\,,\quad a\in\oN\,,\\
        &\kettop{\tfrac{a+2}{p}-1,p;-1}\,,& m=0\,,\quad a\in\oN\,.
      \end{aligned}
    \right.
  \end{equation}
  Similarly,
  \begin{equation}
    \ketbc{m}\tensor\beta_{-\lambda}^a\ketBG{0}\doteq
    \left\{
      \begin{aligned}
        {}&\ket{\tfrac{m-a-2}{p}+1,(a+1)(1+\tfrac{m-2}{p}),p;1-m}\,,
        \quad &
        m\in\oZ\setminus\{1-p\}\,,\quad a\in\oN\,,\\        
        &\kettop{\tfrac{1-a}{p}-1,p;p-1}\,,& m=1-p\,,\quad a\in\oN\,,
      \end{aligned}
    \right.
  \end{equation}
  and also,
  \begin{equation}
    \ketbc{m}\tensor\ketBG{0}\doteq\kettop{\tfrac{m}{p},p;-m}\,.
  \end{equation}
\end{Lemma}

Now, if $\pmV_{\frac{1-r}{p}, p;r-1}(1-r,0)$ were a true Verma module,
we would have singular vectors~\eqref{Eplus} and~\eqref{Eminus},
\begin{align}
  \ket{E(r,1,p)}^{+,r-1}&=
  \cG_{-1}\,\ldots\,\cG_{r-2}\,\kettop{\tfrac{1-r}{p}, p;r-1}\,,
  \label{Eplus-1}\\
  \ket{E(p-r,1,p)}^{-,r-1}&=\cQ_{-p+1}\dots
  \cQ_{-r}\,\kettop{\tfrac{1-r}{p}, p;r-1}\,.
\end{align}
\begin{Lemma}\label{lemma-vanish}
  In the $\N2$ module $\pmV_{\frac{1-r}{p}, p;r-1}(1-r,0)$ generated
  from vector~\eqref{hw-ghost}, we have
  \begin{align}
    \ket{E(r,1,p)}^{+,r-1} &=0\,,\\
    \ket{E(p-r,1,p)}^{-,r-1} &=
    \ketbc{1-p}\tensor\beta_{-\lambda}^{p-r}\,\ketBG{0}\,.
    \label{Eplus-ghost}
  \end{align}
\end{Lemma}
\begin{prf}
  Indeed, we write $\cG_j=\sum_{m\in\oZ+\frac{1}{p}}(m - j
  \lambda)\,\gamma_{j-m}b_m$, then
  \begin{equation*}
    \cG_{r-2}\,\ketbc{1-r}\tensor\ketBG{0}=
    (1-\lambda)(r-1)\,
    b_{r-1-\lambda}\ketbc{1-r}\tensor\gamma_{\lambda-1}\ketBG{0}
  \end{equation*}
  and further, $\cG_{r-3}\,\cG_{r-2}\,\ketbc{1-r}\tensor\ketBG{0}$ is in
  addition proportional to $(1-\lambda)(r-2)$, and so forth;
  therefore, singular vector~\eqref{Eplus-1} vanishes in
  $\pmV_{\frac{1-r}{p}, p;r-1}(1-r,0)$.
\end{prf}

Instead of the vanishing singular vector~\eqref{Eplus-1}, the
$bc\beta\gamma$ representation space contains the vector
\begin{equation}\label{cosingular}
  \begin{split}
    \ket{\cosing}&=b_{-\lambda}\dots b_{r-2-\lambda}b_{r-1-\lambda}\,
    \ketbc{1-r}\tensor\gamma_{\lambda-1}^r\,\ketBG{0}
    =\ketbc{1}\tensor\gamma_{\lambda-1}^r\,\ketBG{0}\\
    &\doteq\ket{\tfrac{r+1}{p},0,p;-1}\,,
  \end{split}
\end{equation}
from which the action of $\cQ$ produces the \hw{} vector: \
$\cQ_{-r+2}\dots \cQ_{0}\cQ_{1}\,\ket{\cosing}=r!\ketbc{1-r}\tensor\ketBG{0}$.
The module $\pmU_{\frac{r+1}{p},0,p;-1}(1,r)$ generated from
$\ket{\cosing}$ has the vanishing charged singular vector $\cQ_{1-r}\dots
\cQ_1\,\ket{\tfrac{r+1}{p},0,p;-1}$. Recalling the submodule
$\pmV_{\frac{r+1}{p}-2,p;p-1}(1-p,r-p)$ generated from singular
vector~\eqref{Eplus-ghost}, we have the mappings
\begin{equation}\label{through}
  \pmV_{\frac{r+1}{p}-2,p;p-1}(1-p,r-p)
  \xrightarrow{[\cQ_{-p+1}\dots \cQ_{-r}]}
  \pmV_{\frac{1-r}{p}, p;r-1}(1-r,0)
  \xrightarrow{[\cQ_{2-r}\dots \cQ_0\cQ_1]}
  \pmU_{\frac{r+1}{p},0,p;-1}(1,r)\,,
\end{equation}
where the square brackets in $\smA\xrightarrow{[\cE]}\smB$ indicate
that the \hw{} vector of $\smA$ is mapped onto the vector obtained by
applying the operator $\cE$ to the \hw{} vector of~$\smB$.  It is
useful to consider the extremal diagram~\cite{[ST4]} describing
relations between the modules involved, see Figs.~\ref{extremal-ghost}
and~\ref{fig:detail}.  The dotted line shows the extremal diagram of
the {\it quotient\/} module
$\pmU_{\frac{r+1}{p},0,p;-1}(1,r)\!\bigm/\!\pmV_{\frac{1-r}{p},
  p;r-1}(1-r,0)$, in which the lower $\circ$ state becomes the
topological singular vector $\ket{E(p-r,2,p)}^{+,-1}$.
\begin{figure}[bt]
  \begin{center}
    \leavevmode\unitlength=.8pt
    \begin{picture}(500,170)
    \qbezier(0,100)(150,200)(300,140)
    \qbezier(300,140)(370,100)(410,0)
    \put(373.5,70){\vector(2,-3){9}}
    \put(381,53){${}_{{}^\bullet}$}  %bottom right
    \put(386,61){${}_{{}^{\ketbc{1-p}\tensor
          \beta_{-\lambda}^{p-r}\,\ketBG{0}}}$}
    \qbezier[1000](53,30)(250,150)(383,55)  %inner parabola
    \put(50,27){${}_{{}^\bullet}$}
    \put(30,13){${}_{{}^\circ}$}
    \put(34,17){\vector(3,2){16}}
    \qbezier(46,0)(70,120)(100,149)
    \put(80,142){\vector(3,1){18}}
    \put(80,140){${}_{{}^\circ}$}    %cosing
    \put(5,150){${}_{{}^{\ketbc{1}\tensor
          \gamma_{\lambda-1}^r\,\ketBG{0}}}$}
    {\linethickness{1pt}
      \qbezier[30](85,139)(160,100)(230,10) %quotient
      }
    \put(299,137){${}_{{}^\bullet}$} %top right
    \put(310,140){${}_{{}^{\ketbc{1-r}\tensor\ketBG{0}}}$}
    \put(97,146){${}_{{}^\bullet}$}  %top left
    \put(103,144){${}_{{}^{\ketbc{0}\tensor
          \gamma_{\lambda-1}^{r-1}\,\ketBG{0}}}$}
  \end{picture}
    \caption{\label{extremal-ghost}{\sf Topological \hw{} states and
        cosingular vectors on the extremal diagram}. Filled dots
      denote the states satisfying twisted topological \hw{}
      conditions. The top parabola is the extremal diagram generated
      from the cosingular vector~$\circ$ (Eq.~\eqref{cosingular}),
      i.e., that of the $\pmU_{\frac{r+1}{p},0,p;-1}(1,r)$ module; the
      vanishing of the charged singular vector
      in~$\pmU_{\frac{r+1}{p},0,p;-1}(1,r)$ results in the cusp at the
      state $\ketbc{1-r}\tensor\ketBG{0}$.  The submodule
      $\pmV_{\frac{1-r}{p}, p;r-1}$ generated from
      $\ketbc{0}\tensor\gamma_{\lambda-1}^{r-1}\ketBG{0}$
      %(or equivalently, from $\ketbc{1-r}\tensor\ketBG{0}$)
      is bounded by the ``vertical'' parabola.  The lower parabola is
      the extremal diagram of the submodule built on singular
      vector~\eqref{Eplus-ghost}.  }
  \end{center}
\end{figure}
\begin{figure}[bth]
  \begin{center}
    \leavevmode \unitlength=1pt
    \begin{picture}(400,180)
      {\linethickness{.8pt}\put(110,100){\vector(0,-1){20}}}
      \put(94,90){${}_{\oQ_0}$}
      \put(50,0){
        {\linethickness{1pt}
          \qbezier[7](120,140)(130,140)(140,140)
          \qbezier[10](140,140)(150,130)(160,120)}
        {\linethickness{.08pt}
          \multiput(95,100)(0,20){5}{\line(1,0){110}}
          \multiput(100,95)(20,0){6}{\line(0,1){90}}
          }
        \put(100,126){${}_{\cQ_2}$}
        \put(100,100){\vector(1,2){20}}
        \put(116,152){${}_{\cQ_1}$}
        \put(120,140){\vector(1,1){20}}
        \put(146,166){${}_{\cQ_0}$}
        {\linethickness{.6pt}\put(140,160){\vector(1,0){20}}}
        \put(160,160){\vector(1,-1){20}}
        \put(180,140){\vector(1,-2){20}}
        \put(140,160){\vector(-1,-2){20}}
        \put(124,123){${}_{\cG_{-2}}$}
        \put(-20,-100){
          {\linethickness{.08pt}
            \multiput(95,120)(0,20){4}{\line(1,0){110}}
            \multiput(100,115)(20,0){6}{\line(0,1){70}}
            }
          \put(100,100){\vector(1,2){20}}
          \put(120,140){\vector(1,1){20}}
          {\linethickness{.6pt}\put(140,160){\vector(1,0){20}}}
          \put(160,160){\vector(1,-1){20}}
          \put(180,140){\vector(1,-2){20}}
          \put(140,160){\vector(-1,-2){20}}
          }
        }
    \end{picture}
    \caption{\label{fig:detail}{\sf Extremal diagram near the
        cosingular vector and the mapping by $\oQ_0$}. The upper
      fragment is the top-left corner of Fig.~\ref{extremal-ghost}
      viewed through a magnifying glass. The cosingular vector is
      mapped by $\cQ_1$ into the module $\pmV_{\frac{1-r}{p}, p;r-1}$
      generated by \ldots,~$\cG_{-2}$,~$\cQ_0$,~$\cQ_{-1}$,~\dots. The
      dotted line spanned by $\cQ_0\,\cQ_{-1}\ldots$ is the extremal
      diagram of the corresponding quotient module. The lower diagram
      gives the result of applying~$\oQ_0$: in the target module,
      there is a {\it sub\/}module isomorphic to the quotient module
      shown in the dotted line, while the submodule from the upper
      diagram has vanished.}
  \end{center}
\end{figure}

The idea now is to observe that singular vector \eqref{Eplus-ghost} is
$\oQ_0$-exact,
\begin{equation}\label{idea}
  \ketbc{1-p}\tensor\beta_{-\lambda}^{p-r}\,\ketBG{0}=
  a_{p,r}\,\oQ_0\,\ketbc{2-p}\tensor
  \beta_{-\lambda}^{2p-r-1}\,\ketBG{0}\,,
\end{equation}
where $a_{p,r}^{-1}=(p-r+1)(p-r+2)\dots(2p-r-1)$,\footnote{In what
  follows, we omit similar factors, once we make sure they are
  nonzero.} while vector \eqref{cosingular} is not $\oQ_0$-closed:
\begin{equation}\label{not-closed}
  \oQ_0\,\ketbc{1}\tensor\gamma_{\lambda-1}^r\,\ketBG{0}=
  \ketbc{0}\tensor\gamma_{\lambda-1}^{p+r-1}\,\ketBG{0}\,.
\end{equation}
As regards the \hw{} vector \eqref{hw-ghost}, further, we have
\begin{equation}
  \oQ_{\geq1-r}\,\ketbc{1-r}\tensor\ketBG{0}=0\,,
\end{equation}
in particular $\oQ_{0}\,\ketbc{1-r}\tensor\ketBG{0}=0$. Moreover,
$\ketbc{1-r}\tensor\ketBG{0}$ is in the cohomology of~$\oQ_0$ for
$1\leq r\leq p-1$, because the state
$\ketbc{2-r}\tensor\beta_{-\lambda}^{p-1}\ketBG{0}$, which is the
candidate for the $\oQ_0$-primitive of $\ketbc{1-r}\tensor\ketBG{0}$
according to the grading, would actually give the desired result only
for $r=p$, which is outside the range of $r$ for the {\it unitary\/}
representations.  In the cohomology of $\oQ_0$, the singular vector
$\ketbc{1-p}\tensor\beta_{-\lambda}^{p-r}\,\ketBG{0}$ in
Fig.~\ref{extremal-ghost} vanishes.  Since the cosingular vector,
further, is {\it not\/} in the cohomology of $\oQ_0$, this suggests
that the cohomology is given precisely by the {\it unitary
  representation\/} $\mK_{r,p;r-1}$, which in terms of the {\it
  Verma\/} modules is given by~\eqref{Verma-quotient}.

We now make this argument more precise.  Informally speaking, we have
to show that although the submodule structure changes in the
free-field realisation, the cohomology of $\oQ_0$ ``effectively''
performs the factorization similar to~\eqref{Verma-quotient}, and that
there are no unwanted submodules in the cohomology. As regards the
latter problem, it is easiest to carry over to the $\N2$ setting the
known results for the $\tSL2$ modules in the Wakimoto bosonisation.
We do this in the next subsection, and then return to the $\N2$
analysis in Sec.~\ref{sec:proof}.

\subsection{Mapping to the Wakimoto bosonisation\label{subsec:to-Wak}}
We recall that the $\tSL2$ currents can be constructed~\cite{[FST]}
from the $\N2$ generators and a bosonic scalar field $\chi$ such that
$\d\chi(z)\d\chi(w)=\frac{-1}{(z-w)^2}$:
\begin{equation}\label{to-sl2}
  J^+= \cQ\,e^\chi,\qquad
  J^0=-\tfrac{p}{2} \cH + \tfrac{p-2}{2}\d\chi\,,\qquad
  J^-=p\, \cG\,e^{-\chi}.
\end{equation}
These $\tSL2_{p-2}$ generators commute with the current
$I^-=\sqrt{\frac{p}{2}}( \cH-\d\chi)$.  Applying this construction to
the $bc\beta\gamma$ realisation of the $\N2$ algebra, we bosonise the
fermionic ghosts as
\begin{equation}\label{bc-ghosts}
  b=e^{-\varphi}\,,\qquad c=e^{\varphi}\,,
\end{equation}
and define a new first-order bosonic system and an independent current
as 
\begin{align}\label{bbeta}
  \bbeta={}&\beta\,e^{\chi+\varphi},\displaybreak[0]\\
  \ggamma={}&\gamma\,e^{-\chi - \varphi},\displaybreak[0]\\
  \sqrt{\tfrac{p}{2}}\,\d\pphi={}&-\tfrac{1}{2}\bbeta\ggamma +
  \tfrac{p+1}{2}\d\varphi + \tfrac{p}{2}\d\chi\,.\label{pphi}
\end{align}
In terms of these fields, $\tSL2_{p-2}$ currents~\eqref{to-sl2} take
the Wakimoto form
\begin{equation}\label{Wakimoto}
  \begin{split}
    J^+ ={}& -\bbeta\,,\\
    J^0 ={}& \sqrt{p/2}\,\d\pphi  + \bbeta\ggamma\,,\\
    J^- ={}& \bbeta\ggamma\ggamma +
    \sqrt{2p}\,\ggamma\d\pphi + (p-2)\d\ggamma\,.
  \end{split}
\end{equation}

Now, further bosonising the $\beta\gamma$ fields from~\eqref{bcBG} as
\begin{equation}\label{beta-gamma}
  \beta=\d\xi\,e^{-\phi},\qquad\gamma=\eta\,e^{\phi},
\end{equation}
where $\eta\xi$ is a first-order fermionic system, we can construct
the bosonic screening for the $\N2$ generators~\eqref{bcBG}:
\begin{equation}\label{Wak-screening}
  S_{\mathrm{W}}=
  \tfrac{1}{2\pi i}\oint\beta\,e^{\frac{1}{p}(\phi-\varphi)}=
  \tfrac{1}{2\pi i}\oint\d\xi\,e^{(\frac{1}{p}-1)\phi-
    \frac{1}{p}\varphi}.
\end{equation}
In terms of the new fields introduced in~\eqref{bbeta}--\eqref{pphi},
this becomes the standard Wakimoto bosonisation screening (i.e., the
one involved in the construction of the Felder-type
resolution~\cite{[BF]}):
\begin{equation}
  S_{\mathrm{W}}=\tfrac{1}{2\pi i}\oint
  \bbeta\,e^{-\sqrt{2/p}\,\pphi}.
\end{equation}
Note that the $\bbeta\ggamma$ system pertaining to the Wakimoto
representation is bosonised as $\bbeta=\d\xi\,e^{-\boldsymbol{\phi}}$,
$\ggamma=\eta\,e^{\boldsymbol{\phi}}$ with the same $\eta$ and $\xi$
as in~\eqref{beta-gamma}.

In terms of the Wakimoto bosonisation ingredients, the fermionic
screening $\oQ_0$ becomes
\begin{equation}\label{QQ0-sl2}
  \oQ_0=\tfrac{1}{2\pi i}\oint e^{\sqrt{2p}\,\pphi}\,\ggamma^{p-1}
  \,e^{\boldsymbol{\phi}}\,.
\end{equation}
This involves the ``constituent'' $\boldsymbol{\phi}$ field in
addition to the $\bbeta\ggamma$ fields as such.  The complete
bosonisation also gives rise to the second fermionic screening
\begin{equation}
  \tfrac{1}{2\pi i}\oint\eta
\end{equation}
that reduces the space of states by eliminating the $\xi_0$
mode~\cite{[FMS]}.  The most important point about the $\oQ_0$
screening in the $\tSL2$ language is that it maps between $\tSL2$
modules with different {\it twists\/} (i.e., the modules transformed
by the spectral flow~\cite{[FST],[FSST]}).  This occurs because the
$e^{\boldsymbol{\phi}}$ factor in~\eqref{QQ0-sl2} changes the
$\bbeta\ggamma$ picture and hence the twist of the $\tSL2$ \hw{}
vector.  Therefore, when applying the recipe of~\cite{[FST]} to
rewrite~\eqref{complex-top} in terms of $\tSL2$ modules in the
Wakimoto bosonisation, we obtain the resolution that ``intersects''
with the one known from~\cite{[BF],[FFr]} at only one point, the
``central'' module, since all the other modules in the $\tSL2$
analogue of~\eqref{complex-top} have non-zero twists.  To avoid
possible misunderstanding regarding how a {\it screening\/} changes
the \hw{} conditions, we show in the next diagram several mappings,
each of which is a counterpart of the one in Fig.~\ref{fig:detail}
(with the extremal diagram conventions as in~\cite{[FST]}; we also
indicate the vertex operators corresponding to the \hw{} vectors of
the ``untwisted'' module (the one carrying the cohomology) and its
submodule):
\begin{equation}
  \unitlength=1.2pt
  \begin{picture}(400,70)
    \put(40,-10){
      \put(24,72){${}_{e^{\frac{r-1}{\sqrt{2p}}\pphi}}$}
      \put(44,34){${}_{\bbeta^{p-r}e^{\frac{r-1}{\sqrt{2p}}\pphi}}$}
      \put(30,89){${}_{J^+_{0}}$}
      \put(-10,80){\line(1,0){50}}
      {\thicklines\put(28,80){\vector(1,0){13}}}
      \put(40,80){\line(1,-1){60}}
      \put(80,50){${}_{J^+_{-1}}$}
      {\thicklines\put(70.5,50){\vector(1,-1){10}}}
      \put(10,40){\line(1,0){70}}
      \multiput(100,40)(130,0){2}{\thicklines
        \put(10,6){${}_{\oQ_{0}}$}
        \put(20,0){\vector(-1,0){15}}
        }
      \put(30,0){
        \put(140,50){${}_{J^+_{0}}$}
        {\thicklines\put(137,40){\vector(1,0){13}}}
        \put(110,40){\line(1,0){80}}
        \put(120,10){\line(1,1){30}}
        {\thicklines\qbezier[10](137,40)(147,30)(157,20)}
        }
      \put(150,0){
        \put(130,50){${}_{J^+_{1}}$}
        {\thicklines\put(140,40){\vector(1,1){10}}}
        \put(110,10){\line(1,1){65}}
        {\thicklines\qbezier[10](140,40)(160,40)(170,40)}
        \put(150,50){\line(-1,-2){20}}
        }
      }
  \end{picture}
\end{equation}
In the central and in the right module, a submodule is in the kernel
of $\oQ_0$; the corresponding quotient module, whose extremal diagram
is shown in the dotted line, is then mapped onto a submodule in the
next module on the left:
\begin{multline}
  \bbeta^{p-r}\,e^{\frac{r-1}{\sqrt{2p}}\pphi}=\oQ_0\,
  \bbeta^{2p-r-1}\,e^{-\boldsymbol{\phi}}\,
  e^{\frac{r-2p-1}{\sqrt{2p}}\pphi},\\
  J^+_0\,\bbeta^{2p-r-1}\,e^{-\boldsymbol{\phi}}\,
  e^{\frac{r-2p-1}{\sqrt{2p}}\pphi} =
  \bbeta^{2p-r}\,e^{-\boldsymbol{\phi}}\,
  e^{\frac{r-2p-1}{\sqrt{2p}}\pphi}=\oQ_0\,
  \bbeta^{3p-r-1}\,e^{-2\boldsymbol{\phi}}\,
  e^{\frac{r-4p-1}{\sqrt{2p}}\pphi},\\
  J^+_{1}\,\bbeta^{3p-r-1}\,e^{-2\boldsymbol{\phi}}\,
  e^{\frac{r-4p-1}{\sqrt{2p}}\pphi}=\oQ_0\,(\dots)\,.
\end{multline}
This pattern continues further right, where there is a sequence of
Wakimoto modules with growing twists (i.e., transformed by the $\tSL2$
spectral flow with $\theta=1,2,\dots$).  Thus, recalling the results
of~\cite{[BF],[FFr]} on the Wakimoto modules, we already know how the
irreducible subquotients arrange in the $\N2$ modules in
$\mG_{n,\nu,p}$.\footnote{\label{foot:aside}As an aside, it would be
  interesting to construct, similarly to~\cite{[FFr]}, the modules
  ``interpolating'' between the Verma modules entering the BGG
  resolution and the {\it twisted\/} Wakimoto modules from the $\tSL2$
  counterpart of~\eqref{complex-top}.}  We now work out some details
in the $\N2$ language.
%and recall the $\tSL2$ counterpart only to save the proof of some
%details regarding the submodule structure.

\subsection{End of the proof of Theorem~\ref{thm:Q0}\label{sec:proof}}
We continue working with the $\nu=0$ case, Eq.~\eqref{complex}. We
observe that there are the following mappings that extend~\eqref{idea}
(up to some nonzero factors, which can be easily restored):
\begin{equation}\label{diagram-2}
  \begin{array}{ccccccc}
    &&\ketbc{2-p}\tensor\beta_{-\lambda}^{3p-r-1}\,\ketBG{0}&
    \!\!\!\xrightarrow{\cQ_{1-p}}\!\!&\ldots\\
    &&\Bigm\downarrow\oQ_0\\
    \ketbc{2-p}\tensor\beta_{-\lambda}^{2p-r-1}\,\ketBG{0}&
    \!\!\!\xrightarrow{\cQ_{1-p}}\!\!&
    \ketbc{1-p}\tensor\beta_{-\lambda}^{2p-r}\,\ketBG{0}\\
    \Bigm\downarrow\oQ_0\\
    \ketbc{1-p}\tensor\beta_{-\lambda}^{p-r}\,\ketBG{0}\,.
    \rule{0pt}{14pt}
  \end{array}
\end{equation}
Evaluating \hw{} conditions for the thus arising states
$\ketbc{2-p}\tensor\beta_{-\lambda}^{mp-r-1}\ketBG{0}$ and
$\ketbc{1-p}\tensor\beta_{-\lambda}^{mp-r}\ketBG{0}$, we find
\begin{align}\label{massive-work}
  \ketbc{1-p}\tensor\beta_{-\lambda}^{mp-r}\,\ketBG{0}&\doteq
  \kettop{\tfrac{r+1}{p} - m - 1,p;p-1}\,,\qquad m=1,2,\dots
  \displaybreak[0]\\
  \ketbc{2-p}\tensor\beta_{-\lambda}^{mp-r-1}\,\ketBG{0}&\doteq
  \ket{\tfrac{r+1}{p} - m,0,p;p-1}\,,%\qquad m=2,3,\dots
  \label{work00}
  \\
  &=\cQ_{2-p}\dots \cQ_{(m-1)p-r}\,\ketbc{(m-1)p-r+1}\tensor
  \ketBG{0}\,,\notag
\end{align}
where, further, $\ketbc{(m-1)p-r+1}\tensor\ketBG{0}\doteq
\kettop{\frac{1-r}{p}+m-1,p;r-1-(m-1)p}$.  For every $m\in\oN$,
therefore, we have a pair of modules
\begin{equation}\label{pair}
  \pmV_{\frac{1-r}{p}+m-1,p;r-1-(m-1)p}((m-1)p-r+1,0)
  \supset\pmV_{\frac{r+1}{p} - m - 1,p;p-1}(1-p,r-mp)
\end{equation}
(of course, the {\it modules\/} are mapped in the opposite directions
to the ${}\xrightarrow{\cQ_{1-p}}{}$ arrows in~\eqref{diagram-2}).  
%We are interested in further mappings involving these modules.
The topological {\it Verma\/} module
$\mV_{\frac{1-r}{p}+m-1,p;r-1-(m-1)p}$ corresponding to the module on
the left-hand side of~\eqref{pair} has singular vectors $\ket{E(p m -
  r,1,p)}^{-,r-1-(m-1)p}$ and $\ket{E(r,m,p)}^{+,r-1-(m-1)p}$. The
first of these corresponds to the twisted topological Verma
module~$\mV_{\frac{r+1}{p} - m - 1,p;p-1}$, and in the ghost
realisation, indeed, evaluates as the \hw{} vector of the submodule on
the right-hand side of~\eqref{pair}.  On the other hand,
$\ket{E(r,m,p)}^{+,r-1-(m-1)p}$ vanishes
in~$\pmV_{\frac{1-r}{p}+m-1,p;r-1-(m-1)p}((m-1)p-r+1,0)$.

Next, in the twisted topological Verma module built on the \hw{} state
as on the right-hand side of~\eqref{pair}, there are singular vectors
$\ket{E((m+1)p-r,1,p)}^{+,p-1}$ and $\ket{E(r,m+1,p)}^{-,p-1}$.  As
regards the first one, we immediately see from~\eqref{Eplus} that it
vanishes in the ghost realisation.  To see what happens to the other
one, we use the recursive nature of Eqs.~\eqref{Tminus}
and~\eqref{Tplus}. In $\pmV_{\frac{r+1}{p} - m - 1,p;p-1}(1-p,r-mp)$,
singular vector~\eqref{Tminus} with the twist parameter~$\theta=p-1$
becomes
\begin{multline}\label{actual-E-minus}
  \ket{E(r,m+1,p)}^{-,p-1}=
  q(1-r-p, (m-1)p)\,\cE^{+,r-(m-1)p-1}(r,m,p)\cdot{}\\
  {}\cdot q((m-1) p - r + 1, -p)\,
  \ketbc{1-p}\tensor\beta_{-\lambda}^{mp-r}\ketBG{0}\,,
\end{multline}
where we can {\it directly evaluate\/} the action of the continued
operator $q((m-1) p - r + 1, -p)$ on the \hw{} state in the ghost
realisation:
\begin{equation}
  q((m-1) p - r + 1, -p)\,
  \ketbc{1-p}\tensor\beta_{-\lambda}^{mp-r}\ketBG{0}=
  \ketbc{(m-1)p-r+1}\tensor\ketBG{0}\,.
\end{equation}
It is essential here that the rest of the right-hand side
of~\eqref{actual-E-minus} contains only the operators from the
universal enveloping of the $\N2$ algebra, since $q(1-r-p, (m-1)p)=
\cQ_{1-r-p}\dots \cQ_{(m-1)p}$ in~\eqref{actual-E-minus}. Therefore,
the singular vector in~\eqref{actual-E-minus} is a descendant of
$\ket{E(r,m,p)}^{+,r-1-(m-1)p}$ and, thus, vanishes.

Continuing in this way, we obtain the two central diagrams (for $m=1$
and~2) and the subsequent diagrams on the right in the sequence
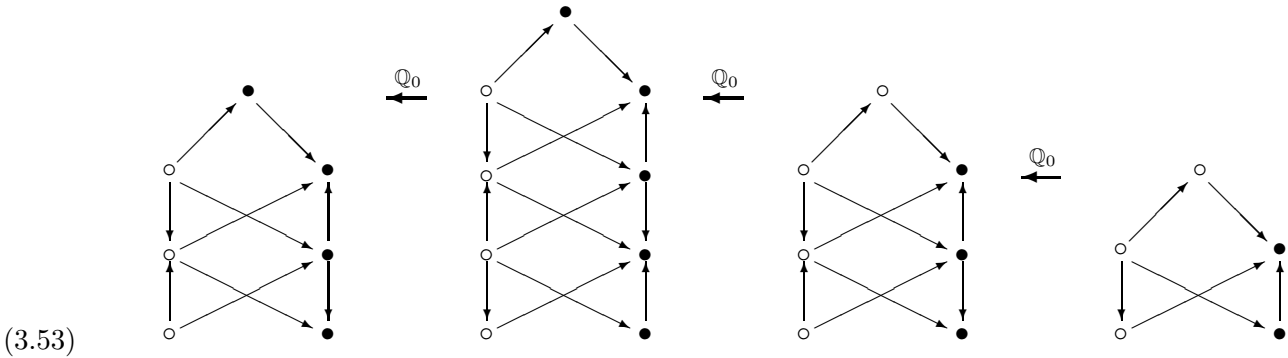
\begin{equation}\label{pattern-first}
  \unitlength=1pt
  \begin{picture}(400,130)
    \put(0,-45){
      \multiput(0,0)(120,30){2}{
        %% the top:
        \put(30,140){$\bullet$}
        \put(6,116){\vector(1,1){22}}
        \put(0,110){$\circ$}
        \put(36,138){\vector(1,-1){22}}
        \put(60,110){$\bullet$}
        \multiput(0,0)(0,-30){2}{          
          \put(7,108){\vector(2,-1){50}}  % \
          \put(7,83){\vector(2,1){50}}  % /
          \put(0,78){$\circ$}
          \put(60,78){$\bullet$}
          }
        \put(3,108){\vector(0,-1){22}}  %down
        \put(63,86){\vector(0,1){22}}   %up
        \put(0,-30){
          \put(63,108){\vector(0,-1){22}}  %down
          \put(3,86){\vector(0,1){22}}   %up
          }
        }
      \multiput(240,0)(120,-30){2}{
        %% the top:
        \put(30,140){$\circ$}
        \put(6,116){\vector(1,1){22}}
        \put(0,110){$\circ$}
        \put(36,138){\vector(1,-1){22}}
        \put(60,110){$\bullet$}
        \multiput(0,0)(0,-30){1}{
          \put(3,108){\vector(0,-1){22}}  %down
          \put(63,86){\vector(0,1){22}}   %up
          \put(7,108){\vector(2,-1){50}}  % \
          \put(7,83){\vector(2,1){50}}  % /
          \put(0,78){$\circ$}
          \put(60,78){$\bullet$}
          }
        }
      \multiput(120,-30)(120,0){2}{        
        \put(7,108){\vector(2,-1){50}}  % \
        \put(7,83){\vector(2,1){50}}  % /
        \put(0,78){$\circ$}
        \put(60,78){$\bullet$}
        }
      \put(120,-30){
        \put(3,108){\vector(0,-1){22}}  %down
        \put(63,86){\vector(0,1){22}}   %up
        }
      \put(240,-30){
        \put(63,108){\vector(0,-1){22}}  %down
        \put(3,86){\vector(0,1){22}}   %up
        }
      \multiput(100,140)(120,0){2}{
        \put(-12,7){${}_{\oQ_0}$}
        \thicklines\put(0,0){\vector(-1,0){15}}
        }
      \put(340,110){
        \put(-12,7){${}_{\oQ_0}$}
        \thicklines\put(0,0){\vector(-1,0){15}}
        }
      }
  \end{picture}
\end{equation}
which is continued %farther right similarly to the above, and 
to the left as we explain momentarily.  The {\it submodule\/}
represented by the filled dots in each diagram is the kernel
of~$\oQ_0$.  These diagrams show the ``adjacency'' of irreducible
subquotients, as discussed after Eq.~\eqref{standard-emb}.
%In particular the diagram in Eq.~\eqref{standard-emb} becomes
%\begin{equation}\label{perverted-emb}
%  \unitlength=1pt
%  \begin{picture}(400,70)
%    \put(0,0){
%      \put(40,0){
%        \put(0,30){$\bullet$}
%        \put(29,6){\vector(-1,1){23}}
%        \put(6,36){\vector(1,1){23}}
%        \put(30,60){$\bullet$}
%        \multiput(30,0)(60,0){3}{
%          \put(0,0){$\bullet$}
%          \put(0,60){$\bullet$}
%          \put(57,7){\vector(-1,1){50}}
%          \put(7,7){\vector(1,1){50}}
%          }
%        \put(37,3){\vector(1,0){50}}
%        \put(97,63){\vector(1,0){50}}
%        \put(87,63){\vector(-1,0){50}}
%        \put(147,3){\vector(-1,0){50}}
%        \put(207,63){\vector(-1,0){50}}
%        \put(157,3){\vector(1,0){50}}
%        \multiput(220,3)(170,0){1}{\Large\dots}
%        \multiput(220,61)(170,0){1}{\Large\dots}
%        }
%      }
%  \end{picture}
%\end{equation}

%\medskip

Now, %diagram~\eqref{pattern-first}
this can be continued to the left of the center as follows.  The
module $\pmV_{\frac{1-r}{p}, p;r-1}(1-r,0)$ can equivalently be
written as $\pmV_{\frac{r+1}{p}-1, p;-1}(0,r-1)$, which simply
corresponds to taking the next cusp to the left on the extremal
diagram as the \hw{} vector (see Fig.~\ref{extremal-ghost})
$\ketbc{0}\tensor\gamma_{\lambda-1}^{r-1}\,\ketBG{0}\doteq
\kettop{\tfrac{r+1}{p}-1,p;-1}$.  As we have seen,
$\cQ_1\,\ketbc{1}\tensor\gamma_{\lambda-1}^r\,\ketBG{0}
=\ketbc{0}\tensor\gamma_{\lambda-1}^{r-1}\,\ketBG{0}$.  This and
Eq.~\eqref{not-closed} can now be continued as follows (again, up to
nonvanishing factors):
\begin{equation}\label{diagram-1}
  \begin{array}{ccccccc}
    &&&&\ketbc{1}\tensor\gamma_{\lambda-1}^r\,\ketBG{0}&
    \!\!\!\xrightarrow{\cQ_1}\!\!&
    \ketbc{0}\tensor\gamma_{\lambda-1}^{r-1}\,\ketBG{0}\\
    &&&&\Bigm\downarrow\oQ_0\\
    &&\ketbc{1}\tensor\gamma_{\lambda-1}^{p+r}\,\ketBG{0}&
    \!\!\!\xrightarrow{\cQ_1}\!\!&
    \ketbc{0}\tensor\gamma_{\lambda-1}^{p+r-1}\,\ketBG{0}\\
    &&\Bigm\downarrow\oQ_0\\
    \dots&
    \!\!\xrightarrow{\cQ_1}\!\!&
    \ketbc{0}\tensor\gamma_{\lambda-1}^{2p+r-1}\,\ketBG{0}
  \end{array}
\end{equation}
We thus obtain the ``gravitational descendants'' of the \hw{}
state,\footnote{That the above diagram involves the mode $\cQ_1$ is in
  accordance with the spectral flow transform by $\theta=-1$ of \hw{}
  states~\eqref{replicas}; twist~0 would correspond to~$\cQ_0$, and to
  $\cQ_{-p}$ in~\eqref{diagram-2}.  Note also that by restoring the
  coefficients, we obtain~\eqref{tic-tac-toe}.}
\begin{equation}\label{replicas}
  \ketbc{0}\tensor\gamma_{\lambda-1}^{mp+r-1}\,\ketBG{0}\doteq
  \kettop{\tfrac{r+1}{p}+m-1,p;-1}\,,\quad m=0,1,2\dots
\end{equation}
{}From the twisted topological \hw{} conditions satisfied by these
vectors, it follows, in particular, that the
arrows~${}\xrightarrow{\cQ_1}{}$ cannot be inverted:
$\cG_{-1}\,\ketbc{0}\tensor\gamma_{\lambda-1}^{mp+r-1}=0$.  
%Further, writing
%$\tfrac{r+1}{p}+m-1=\hminus(r+pm,1,p)=\hplus(p-r,m+1,p)$, we see in
%accordance with the formulae of the Appendix which singular vectors
%can be expected in the modules built on \hw{} states~\eqref{replicas}.

%On the other hand, 
In each of the modules $\pmU_{\frac{r+1}{p}+m,0,p;-1}$ generated from
the states $\ketbc{1}\tensor\gamma_{\lambda-1}^{pm+r}\,\ketBG{0}\doteq
\ket{\tfrac{r+1}{p}+m,0,p;-1}$ in~\eqref{diagram-1}, the charged
singular vector~$\cQ_{1-r-pm}\dots
\cQ_1\,\ketbc{1}\tensor\gamma_{\lambda-1}^{pm+r}\,\ketBG{0}$ vanishes.
Each of these modules also has the charged singular vector
$\cQ_1\,\ket{\frac{r+1}{p}+m,0,p;-1}$, the {\it quotient\/} over which
is isomorphic to $\pmV_{\frac{r+1}{p}+m,p;-1}$; on the other hand, the
{\it submodule\/} generated from the $\cQ_1$-singular vector in the
next module $\pmU_{\frac{r+1}{p}+m+1,0,p;-1}$ is isomorphic
to~$\pmV_{\frac{r+1}{p}+m,p;-1}$ (see~Fig.~\ref{fig:detail}).
Continuing in this way, we also recall that diagrams~\eqref{diagram-1}
and~\eqref{diagram-2} are glued together as shown in
Eq.~\eqref{through},
$\ketbc{0}\tensor\gamma_{\lambda-1}^{r-1}\,\ketBG{0}
\xrightarrow{\cQ_{1-p}\dots \cQ_{-r}\cQ_{2-r}\dots \cQ_0}
\ketbc{1-p}\tensor\beta_{-\lambda}^{p-r}\,\ketBG{0}$.

{}From the the ``adjacency'' structure of the irreducible
subquotients, as described in~\eqref{pattern-first}, we immediately
conclude that the cohomology is concentrated at one term and that it
is the unitary representation~$\mK_{r,p}$.\hfill
\mbox{\rule{.5em}{.5em}}

\section{The ``symmetric'' realisation\label{sec:symmetric}}
\subsection{Generalities} We now consider the realisation of the $\N2$
algebra~\cite{[MSS],[OS],[Ito]} that has the following form in the
conventions corresponding to~\eqref{topalgebra}:
\begin{equation}\label{zeta-untwisted}
  \begin{split}
    \cG(z)&=C(z)A(z) - \d C(z)\,,\\
    \cQ(z)&=B(z)\bar A(z) - \tfrac{1}{p}\d B(z)\,,\\
    \cH(z)&=\bar A(z) -
    \tfrac{1}{p}A(z) - B(z)C(z)\,,\\
    \cT(z)&=\d B(z)\,C(z) + \bar A(z)A(z) -
    \d\bar A(z)\,,
  \end{split}
\end{equation}
where the free-field operator products are
\begin{equation}
  \bar A(z)\,A(w) = \frac{1}{(z-w)^2}\,,\quad
  B(z)\,C(w) = \frac{1}{z-w}\,.
\end{equation}
We call this realisation {\it symmetric\/} because the $\cG$ and $\cQ$
operators are expressed through the free fields in an essentially
symmetric way, which is in contrast with the (inequivalent)
``asymmetric'' realisation~\cite{[GS2],[BLNW]} of the $\N2$ algebra.
We expand into modes as $A(z)=\sum_{n\in\oZ}A_nz^{-n-1}$, 
$\bar A(z)=\sum_{n\in\oZ}\bar A_nz^{-n-1}$, and 
$B(z)=\sum_{n\in\oZ}B_nz^{-n}$, $C(z)=\sum_{n\in\oZ}C_nz^{-n-1}$.
\begin{Lemma}
  The $\N2$ spectral flow is induced by transforming the free fields
  entering~\eqref{zeta-untwisted} as follows:
  \begin{equation}\label{BC-spectral}
    \begin{aligned}
      C_n&{}\mapsto C_{n+\theta}\,,&\qquad
      &A_n&\mapsto A_n + \theta\delta_{n,0}\,,\\
      B_n&{}\mapsto B_{n-\theta}\,,&
      &\bar A_n&\mapsto\bar A_n -
      \tfrac{\theta}{p}\delta_{n,0}\,.\\
    \end{aligned}
  \end{equation}
\end{Lemma}
\begin{prf} %This is straightforward;
  One only has to note that under this transformation, the composite
  $BC$ operators acquire, in the standard way, a contribution coming
  from normal reordering (which is omitted from the notations), %e.g.,
  \begin{equation}
    (BC)_n\mapsto(BC)_n - \theta\delta_{n,0}\,,\qquad
    (\d B\,C)_n\mapsto
    (\d B\,C)_n -\theta(BC)_n +
    \tfrac{\theta^2+\theta}{2}\delta_{n,0}\,.
  \end{equation}
\end{prf}
There are two fermionic screenings
\begin{equation}\label{two-fermionic}
  \SB=\tfrac{1}{2\pi i}\oint Be^{X}\,,\qquad
  \SC=\tfrac{1}{2\pi i}\oint Ce^{p\bar X}\,,
\end{equation}
where \ $\d X = A$ \ and \ $\d\bar X = \bar A$. \ Note that,
obviously, the screenings are unchanged under~\eqref{BC-spectral}.

We now introduce modules over these free fields.
$\Lambda=\Lambda_0(0)$ is the $BC$ module generated from $\ketBC{0}$
(see \eqref{bcannihil} and~\eqref{bceigen}, where now $\lambda=0$);
next, let the state $\ketAA{\bar a, a}$ be such that
\begin{gather}
  \bar A_0\ketAA{\bar a, a}=\bar a\ketAA{\bar a, a}\,,\quad
  A_0\ketAA{\bar a, a}=a\ketAA{\bar a, a}\,,\\
  \bar A_{\geq1}\ketAA{\bar a, a}=A_{\geq1}\ketAA{\bar a, a}=0
\end{gather}
and let $\mH_{\bar a,a}$ be the Fock module generated from
$\ketAA{\bar a, a}$. We define the ghost representation space
\begin{equation}
  \oG_{r,p}=\Lambda\tensor\bigoplus_{m,n\in\oZ}\mH_{n,mp+r-1}\,.
\end{equation}
It is in this space that we will identify the unitary
representation~$\mK_{r,p;0}$. The other~$\mK_{r,p;\theta}$ then follow
by applying the spectral flow transform in accordance
with~\eqref{BC-spectral}.

\medskip

\noindent{\it Notation for the $\N2$ modules}. Similarly to the
conventions used in Sec.~\ref{sec:ghost-structure}, we denote the
module generated from $\ketBC{M}\tensor\ketAA{\bar a, a}\in\oG_{r,p}$
by $\dmU_{h,\ell,p;\theta}(M,\bar a,a)$ whenever
$\ketBC{M}\tensor\ketAA{\bar a, a}\doteq\ket{h,\ell,p;\theta}$ (i.e.,
the state satisfies twisted massive \hw{}
conditions~\eqref{verma}--\eqref{masshw-2}), and by
$\dmV_{h,p;\theta}(M,\bar a,a)$ in the case where
$\ketBC{M}\tensor\ketAA{\bar a, a}\doteq\kettop{h,p;\theta}$.  We will
sometimes omit the $(M,\bar a,a)$ arguments, explicitly indicating
instead the vector(s) from which the module is generated.  Note,
however, that there will also appear $\N2$ modules that are not
generated from a single vector.

\medskip

In $\oG_{r,p}$, we take the state $\ket{0}_{BC}\tensor\ketAA{0,r-1}$
which satisfies the (twist-zero) topological \hw{} conditions with
respect to $\N2$ generators~\eqref{zeta-untwisted}:
\begin{equation}\label{central-hw}
  \ket{0}_{BC}\tensor\ketAA{0,r-1}
  \doteq\kettop{\tfrac{1-r}{p},p}\,.
\end{equation}

\subsection{The butterfly resolution} Let
$\dmV_{\frac{1-r}{p},p}(0,0,r-1)$ be the $\N2$ module generated
from~\eqref{central-hw}.  It is a submodule in the ``central'' term
\textcircled{\small$\bullet$} (the one carrying the cohomology) in the
{\it butterfly resolution\/}:\footnote{Or rather a {\it sand-glass},
  if rotated by $135^\circ$.}
\begin{equation}
  \label{butterfly}
  \unitlength=1pt
  \begin{picture}(400,250)
    \put(-10,20){
      \put(192.9,116){\textcircled{\phantom{\small$\bullet$}}}
      \multiput(126,214)(35,0){3}{$\vdots$}
      \multiput(82,118)(0,32){3}{$\dots$}
      \multiput(168,180)(0,-32){3}{
        \multiput(0,0)(-35,0){3}{
          \put(27.5,0){$\bullet$}\put(4,0){$\xleftarrow{\SC}{}$}
          \put(30,8){\vector(0,1){20}}
          \put(32,20){${}_{\SB}$}
          }
        }
      \put(228,82){
        %\put(-1.8,0){\textcircled{\phantom{\small$\bullet$}}}
        \put(-5,12){\vector(-1,1){18}}
        }
      \multiput(224,87)(0,-32){3}{
        \multiput(0,0)(35,0){3}{
          \put(0,0){$\bullet\xleftarrow{\SC}{}$}
          \put(3,-24){\vector(0,1){20}}
          \put(6,-10){${}_{\SB}$}
          }
        }
      \multiput(225,-16)(35,0){3}{$\vdots$}
      \multiput(334,89)(0,-32){3}{$\dots$}
      }
  \end{picture}
\end{equation}
On the right wing, the modules are
$\dmU_{\frac{r+1}{p}-n-m,0,p;np-r}(np-r,-m,r-np-1)$, which are labeled
by positive integers $m$ and $n$, where $n$ labels columns (with $n=1$
for the left column) and $m$ labels rows (with $m=1$ for the top row).
On the left wing, the construction is ``dual'' in the obvious sense
(the arrows reversed), the modules are labeled by nonnegative integers
$m$ and $n$, with $m=0$ for the bottom row and $n=0$ for the right
column, the $(m,n)$ module being the result of gluing together the
modules generated from the states
$\ketBC{-r-np}\tensor\ketAA{m,np+r-1}$ and
$\ketBC{mp+1}\tensor\ketAA{m,np+r-1}$.  In particular, the
\textcircled{\small$\bullet$}-module is obtained by gluing together
the $\N2$ modules generated from $\ketBC{-r}\tensor\ketAA{0,r-1}$ and
from $\ketBC{1}\tensor\ketAA{0,r-1}$, which have a common submodule
generated from~\eqref{central-hw}.
\begin{Thm}\label{thm:butterfly}
  Diagram~\eqref{butterfly} consisting of $\N2$ representations
  on~$\oG_{r,p}$ is exact except at the \textcircled{\small$\bullet$}
  module generated from the vectors $\ketBC{-r}\tensor\ketAA{0,r-1}$
  {\em and\/} $\ketBC{1}\tensor\ketAA{0,r-1}$, where the cohomology is
  given by the unitary $\N2$ representation~$\mK_{r,p}$.
\end{Thm}
\begin{Rem}
  We have selected an $r$ from the range $1\leq r\leq p-1$
  arbitrarily, and defined the space $\oG_{r,p}$ depending on $r$. If
  we define
  \begin{equation}
    \oG_{*,p}=\Lambda\tensor\bigoplus_{m,n\in\oZ}\mH_{n,m}\,,
  \end{equation}
  the cohomology on this space is \ $\displaystyle
  \bigoplus_{r=1}^{p-1}\mK_{r,p}$.
\end{Rem}
\begin{Rem}
  The $\N2$ spectral flow acts on the states from $\oG_{*,p}$
  according to
  \begin{equation}
    \ketBC{n}\tensor\ketAA{\bar a, a}\mapsto
    \ketBC{n+\theta}\tensor\ketAA{\bar a+\tfrac{\theta}{p},a-\theta}\,.
  \end{equation}
  This allows us to apply the spectral flow to the data in
  Theorem~\ref{thm:butterfly} (with $\oG_{r,p}$ mapping appropriately)
  in order to obtain all the unitary representations
  $\mK_{r,p;\theta}$ from Sec.~\ref{sec:general}.  Without a loss of
  generality, therefore, we work with a particular twist---and, thus,
  with the above~$\oG_{r,p}$ space---as fixed by the choice made
  in~\eqref{central-hw}.
\end{Rem}
\begin{prf}
  The construction of the resolution %goes in several steps.  It
  is based on charged singular vectors~\eqref{ECh} and on the
  topological singular vectors \eqref{Eplus} and~\eqref{Eminus} (those
  with $s=1$).  In the free-field realisation, these singular vectors
  may vanish, in which case we may find cosingular vectors in the same
  grade; if, on the other hand, the singular vector does not vanish,
  we consider the corresponding submodule and find similar singular
  vectors in it.  Thus, for the state~\eqref{central-hw} and for other
  \hw{} states appearing on the way, we determine their annihilation
  conditions; these would be either twisted topological \hw{}
  conditions~\eqref{annihiltop} or only twisted massive \hw{}
  conditions~\eqref{masshw}.  We then evaluate the eigenvalues of the
  appropriate Cartan generators (using~\eqref{H0-top}, \eqref{L0-top},
  or~\eqref{masshw}), look for the singular vectors, and investigate
  which (sub)modules are in the image or in the kernel of the
  screening operators; the action of the screenings o states
  from~$\oG_{r,p}$ is evaluated via elementary conformal field theory
  calculations.

  \medskip

  \noindent
  \underline{\it The middle}.  We begin with~\eqref{central-hw}
  and the corresponding $\N2$
  module~$\dmV_{\frac{1-r}{p},p}(0,0,r-1)$.  Evaluating the singular
  vector $\ket{E(r,1,p)}^{+}$ (see~\eqref{hplusminus}--\eqref{Eplus})
  in this module, we see that it vanishes.  Because of this vanishing,
  the state $\ketBC{1-r}\tensor\ketAA{0,r-1}$ satisfies twisted {\it
    topological\/} \hw{} conditions:
  \begin{equation}\label{left-corner}
    \ketBC{1-r}\tensor\ketAA{0,r-1}\doteq
    \kettop{\tfrac{r+1}{p}-1,p;-r}\,.
  \end{equation}

  We now observe that vector~\eqref{left-corner} is in the image
  of~$\SB$:
  \begin{equation*}
    \ketBC{1-r}\tensor\ketAA{0,r-1}
    =\SB\,\ketBC{-r}\tensor\ketAA{-1,r-1}\,,
  \end{equation*}
  where we find $\ketBC{-r}\tensor\ketAA{-1,r-1}\doteq
  \ket{\frac{r+1}{p}-1,0,p;-r}$.  In the module
  $\dmU_{\frac{r+1}{p}-1,0,p;-r}$ generated from
  $\ketBC{-r}\tensor\ketAA{-1,r-1}$, there is a charged singular
  vector
  \begin{equation}
    \begin{split}
      \cG_{-p}\dots\cG_{-r-1}\,\ketBC{-r}\tensor\ketAA{-1,r-1}={}&
      \ketBC{-p}\tensor\ketAA{-1,r-1}\\
      ={}&\SC\,\ketBC{1-p}\tensor\ketAA{-1,r-p-1}\,.
    \end{split}
  \end{equation}
  The thus arising vector $\ketBC{1-p}\tensor\ketAA{-1,r-p-1}$
  generates the module $\dmU_{\frac{r+1}{p}-2,0,p;p-r}(p-r,-1,r-p-1)$,
  which is the $(m=1,n=1)$ case of the modules filling the right wing,
  as we describe momentarily.  Note that we could equally well have
  arrived at the same module by first noticing that
  $\ketBC{0}\tensor\ketAA{0,r-1}=
  \SC\,\ketBC{1}\tensor\ketAA{0,r-p-1}$, where, further,
  $\cQ_{r-p}\dots\cQ_{-1}\ketBC{1}\tensor\ketAA{0,r-p-1}=
  \ketBC{p-r+1}\tensor\ketAA{0,r-p-1}=
  \SB\,\ketBC{p-r}\tensor\ketAA{-1,r-p-1}$, with
  $\ketBC{p-r}\tensor\ketAA{-1,r-p-1}$ and
  $\ketBC{1-p}\tensor\ketAA{-1,r-p-1}$ being descendants of each
  other.  This defines the central arrow in~\eqref{butterfly} as
  $\SB\circ \SC$, which equals $\SC\circ \SB$ up to the (irrelevant)
  sign.

  \medskip

  \noindent
  \underline{\it The right wing}.  Assigning $n=1$ to the left column
  and $m=1$ to the top row, the module in column~$n$ and line~$m$ is
  given by
  \begin{multline}\label{equality}
    \dmU_{\frac{r+1}{p}-n-m,0,p;np-r}(np-r,-m,r-np-1)={}\\
    {}=\dmU_{\double{n+m-\frac{r+1}{p}},p;1-mp}(1-mp,-m,r-np-1)\,,
  \end{multline}
  which can be generated from any of the vectors
  \begin{equation*}
    \ketBC{np-r}\tensor\ketAA{-m,r-np-1}\doteq
    \ket{\tfrac{r+1}{p}-m-n,0,p;np-r}
  \end{equation*}
  or
  \begin{equation*}
    \ketBC{1-mp}\tensor\ketAA{-m,r-np-1}\doteq
    \ket{\double{n+m-\tfrac{r+1}{p}},p;1-mp}\,.
  \end{equation*}
  We first note that the equality in~\eqref{equality} occurs in view
  of the relations
  \begin{gather*}
    \cQ_{-np+r+1}\dots\cQ_{mp-1}\,
    \ketBC{1-mp}\tensor\ketAA{-m,r-np-1}=
    \ketBC{np-r}\tensor\ketAA{-m,r-np-1}\\
    \ketBC{1-mp}\tensor\ketAA{-m,r-np-1}=
    \cG_{1-mp}\dots\cG_{np-r-1}\ketBC{np-r}\tensor\ketAA{-m,r-np-1}
  \end{gather*}
  (which hold up to {\it nonvanishing\/} factors), and therefore, the
  states $\ketBC{1-mp}\tensor\ketAA{-m,r-np-1}$ and
  $\ketBC{np-r}\tensor\ketAA{-m,r-np-1}$ generate the same module.

  Next, a simple calculation shows that there are the following
  charged singular vectors in the module in Eq.~\eqref{equality}:
  \begin{equation*}
    \begin{array}{lcccl}
      {}\\[-24pt]
      \cG_{-mp}\,\ketBC{1-mp}\tensor\ketAA{-m,r\!-\!np\!-\!1}
      \kern-6pt&\in&\kern-6pt
      \dmU_{\frac{r+1}{p}\!-\!m\!-\!n,0,p;np\!-\!r}
      \kern-6pt&\ni&\kern-6pt
      \cQ_{r-np}\,\ketBC{np-r}\tensor\ketAA{-m,r-np-1}\\
      ~||&{}&{}&{}&~||\\[2pt]
      \ketBC{-mp}\tensor\ketAA{-m,r-np-1}&{}&{}&{}&
      \kern-10pt
      \ketBC{1+np-r}\tensor\ketAA{-m,r-np-1}\\
      ~||\cdot&{}&{}&{}&~||\cdot\\[2pt]
      \kettop{n+m+\frac{1-r}{p},p;-mp}&{}&{}&{}&
      \kern-10pt
      \kettop{\frac{r+1}{p}-n-m-1,p;np-r}\\
      ~||&{}&{}&{}&~||\\[2pt]
      \SC\,\ketBC{1-mp}\tensor\ketAA{-m,r-(n+1)p-1}\kern-30pt
      &{}&{}&{}&
      \kern-16pt \SB\,\ketBC{np-r}\tensor\ketAA{-m-1,r-np-1}\\[-6pt]
      {}
    \end{array}
  \end{equation*}
  Evaluating the annihilation and eigenvalue conditions satisfied by
  $\ketBC{1-mp}\tensor\ketAA{-m,r-(n+1)p-1}$ and
  $\ketBC{np-r}\tensor\ketAA{-m-1,r-np-1}$, we see that the same steps
  as above apply to the modules generated from these vectors: each of
  these modules has two charged singular vectors, which are precisely
  in the image of one of the screenings. The ``cross'' combination of
  two screenings follows by repeating the above calculations twice
  more. This can be best explained by Fig.~\ref{fig:4}, the outcome
  being that there exist the mappings
  \begin{equation}
    \begin{array}{lclclclc}
      &&\vdots&{}&\vdots\\
      &&\UParrow{\SB}&{}&\UParrow{\SB}\\
      \cdots&\xleftarrow{\SC}&
      \dmU_{\frac{r+1}{p}-n-m,0,p;np-r}&\xleftarrow{\SC}&
      \dmU_{\frac{r+1}{p}-n-m-1,0,p;(n+1)p-r}&\xleftarrow{\SC}&
      \dots\\
      &&\UParrow{\SB}&{}&\UParrow{\SB}\\
      \cdots&\xleftarrow{\SC}&
      \dmU_{\frac{r+1}{p}-n-m-1,0,p;np-r}&\xleftarrow{\SC}&
      \dmU_{\frac{r+1}{p}-n-m-2,0,p;(n+1)p-r}&\xleftarrow{\SC}&
      \dots\\
      &&\UParrow{\SB}&{}&\UParrow{\SB}\\
      &&\vdots&{}&\vdots
    \end{array}
  \end{equation}
  which constitute the pattern filling the right wing of the
  butterfly.  
  \begin{figure}[t]
    \begin{center}
      \leavevmode
      \begin{picture}(480,240)
        \put(20,-70){
        %%11
        \qbezier(-10,242)(60,302)(160,298)
        \put(17,259){${{}_{\ket{1-mp}\ket{-m,r-np-1}}}$}
        \put(11,257){\footnotesize$\bullet$}
        \put(-13,262){${}_{\cG_{-mp}}$}
        {\thicklines\put(12,258){\vector(-3,-2){12}}}
        \qbezier(0,250)(15,252)(32,252)
        \put(50,303){${{}_{\ket{np-r}\ket{-m,r-np-1}}}$}
        \put(132,295.5){\footnotesize$\bullet$}
        \put(134,306){${}_{\cQ_{r-np}}$}
        {\thicklines\put(136,298){\vector(1,0){14}}}
        \qbezier(150,298)(140,295)(130,280)
        \put(149,200){\vector(0,1){30}}
        \put(152,210){\footnotesize$\SB$}
        %%
        %%21
        %%
        \put(15,-160){
          \qbezier(25,242)(60,302)(160,340)
          \put(40,259){${{}_{\ket{1-(m+1)p}\ket{-m-1,r-np-1}}}$}
          \put(35,257){\footnotesize$\bullet$}
          \put(-6,260){${}_{\cG_{-(m+1)p}}$}
          {\thicklines\put(36,258){\vector(-2,-3){8}}}
          \qbezier(27.5,246)(40,250)(60,250)
          \put(38,332){${{}_{\ket{np-r}\ket{-m-1,r-np-1}}}$}
          \put(132,327){\footnotesize$\bullet$}
          \qbezier(150,336)(140,326)(130,310)
            \put(128,341){${}_{\cQ_{r-np}}$}
            {\thicklines\put(135.5,330.5){\vector(3,1){13}}}
          {\thicklines\qbezier[20](133,328.5)(140,332)(170,330)}
          }
        \put(180,253){\footnotesize$\SC$}
        \put(200,250){\vector(-1,0){30}}
        %%
        %%12
        %%
        \put(260,0){\qbezier(-10,243)(60,268)(160,246)
          \put(21,248){${{}_{\ket{1-mp}\ket{-m,r-(n+1)p-1}}}$}
          \put(14,248){\footnotesize$\bullet$}
          \put(-5,258){${}_{\cG_{-mp}}$}
          {\thicklines\put(14,250){\vector(-4,-1){12}}}
          \qbezier(3,247)(15,246)(32,240)
          \put(75,266){${{}_{\ket{(n+1)p-r}\ket{-m,r-(n+1)p-1}}}$}
          \put(132,249){\footnotesize$\bullet$}
          \put(136,258){${}_{\cQ_{r-(n+1)p}}$}
          {\thicklines\put(136,251){\vector(4,-1){12}}}
          \qbezier(148,248)(140,248)(120,240)
          }
        \put(190,90){\footnotesize$\SC$}
        \put(210,86.5){\vector(-1,0){30}}
        \put(408,180){\vector(0,1){30}}
        \put(410,190){\footnotesize$\SB$}
        %%
        %%22
        %%
        \put(273.6,-160){
          \qbezier(-30,236)(60,298)(160,298)
          \put(-7,248){${{}_{\ket{1-(m+1)p}\ket{-m-1,r-(n+1)p-1}}}$}
          \put(-16,244){\footnotesize$\bullet$}
          \put(-53,251){${}_{\cG_{-(m+1)p}}$}
          {\thicklines\put(-15,246){\vector(-3,-2){12}}}
          \qbezier(-26,238)(-10,242)(12,242)
          \put(5,303){${{}_{\ket{(n+1)p-r}\ket{-m-1,r-(n+1)p-1}}}$}
          \put(132,295){\footnotesize$\bullet$}
          \put(134,306){${}_{\cQ_{r-(n+1)p}}$}
          {\thicklines\put(136,297.6){\vector(1,0){13}}}
          \qbezier(150,298)(140,295)(130,280)
          }
        }
      \end{picture}
      \caption{\label{fig:4}{\sf Extremal diagrams of  modules mapped
          by the screening operators}.  For brevity, we omit the
        tensor product sign and the subscripts pertaining to different
        ket-vectors.  Charged singular vectors and the corresponding
        submodules are shown as explained in~\eqref{charged-picture}.
        The filled dots show the states that are {\it mapped into\/}
        the charged singular vectors by the screenings. The dotted
        line shows in one case (and can be similarly drawn in other
        cases) the extremal diagram of the quotient module over the
        corresponding singular vector.  The corresponding submodule
        being in the kernel of the screening, the dotted line is
        mapped onto the submodule extremal diagram in the target
        module.}
    \end{center}
  \end{figure}
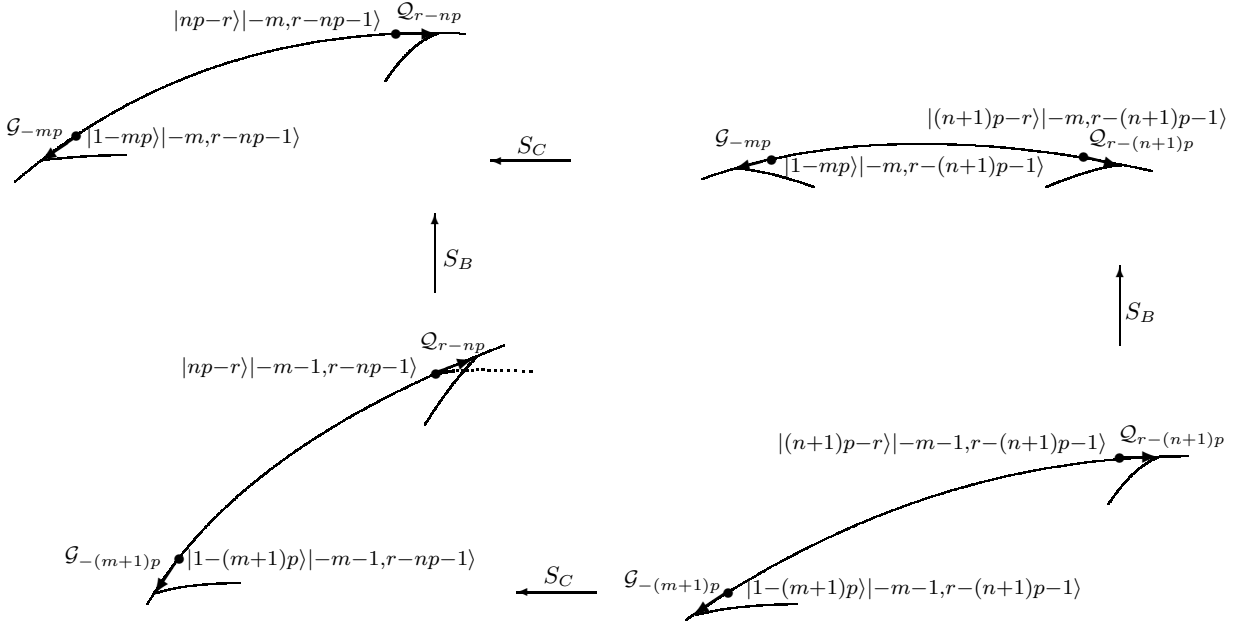

  \medskip

  \noindent
  \underline{\it The left wing}.  We label the modules on the left
  wing by nonnegative integers $m$ and $n$, with $m=0$ for the bottom
  line and $n=0$ for the vertical border.  The $(m,n)$ module is
  obtained by gluing together the modules generated from
  \begin{equation}\label{left-B}
    \ketBC{-r-np}\tensor\ketAA{m,np+r-1}
    \doteq\ket{\tfrac{r+1}{p}+m+n,0,p;-r-np}
  \end{equation}
  and
  \begin{equation}\label{left-C}
    \ketBC{mp+1}\tensor\ketAA{m,np+r-1}\doteq
    \ket{\double{-\tfrac{r+1}{p}-m-n},p;mp+1}\,.
  \end{equation}
  {\it Gluing\/} refers to the fact that in the module generated
  from~\eqref{left-B}, there is a submodule generated from the
  singular vector
  \begin{equation}\label{sing-Q}
    \begin{split}
      \cQ_{r+np}\ketBC{-r-np}\tensor\ketAA{m,np+r-1}
      &=\ketBC{1-r-np}\tensor\ketAA{m,np+r-1}\\
      &\doteq\kettop{\tfrac{r+1}{p}+m+n-1,p;-r-np}\,,
    \end{split}
  \end{equation}
  and {\it the same\/} submodule is also generated from the singular
  vector
  \begin{equation}\label{sing-G}
    \cG_{mp}\ketBC{mp+1}\tensor\ketAA{m,np+r-1}=
    \ketBC{mp}\tensor\ketAA{m,np+r-1}
  \end{equation}
  in the module generated from~\eqref{left-C}, because the states
  $\ketBC{1-r-np}\tensor\ketAA{m,np+r-1}$ and
  $\ketBC{mp}\tensor\ketAA{m,np+r-1}$ are descendants of each other.
  Now, applying the screening operators, we have
  \begin{align}\label{after-B}
    \begin{split}
      \SB\,\ketBC{-r-np}\tensor\ketAA{m,np+r-1}
      &=\ketBC{1-r-np}\tensor\ketAA{m+1,np+r-1}\\
      &=\cQ_{r+np}\,\ketBC{-r-np}\tensor\ketAA{m+1,np+r-1}
    \end{split}
  \end{align}
  and
  \begin{align}\label{after-C}
    \begin{split}
      \SC\,\ketBC{mp+1}\tensor\ketAA{m,np+r-1}
      &=\ketBC{mp}\tensor\ketAA{m,(n+1)p+r+1}\\
      &=\cG_{mp}\ketBC{1+mp}\tensor\ketAA{m,(n+1)p+r+1}\,,
    \end{split}
  \end{align}
  which shows that the pattern~\eqref{left-B}--\eqref{sing-G} is
  reproduced in~\eqref{after-B} and~\eqref{after-C} with $m\mapsto
  m+1$ and $n\mapsto n+1$, respectively.  In the module generated
  from~\eqref{left-B} {\it and\/} \eqref{left-C}, further, $\Ker \SB$
  is generated from~\eqref{left-C}, and $\Ker \SC$
  from~\eqref{left-B}, while $\Ker \SB\cap\Ker \SC$ is the submodule
  generated from~\eqref{sing-Q}.  In this way, the entire left wing is
  filled (in the way that is in the obvious sense dual to the pattern
  on the right wing).
  
  We now recall the standard fact that each horizontal or vertical
  sequence of mappings by {\it one\/} fermionic screening represented
  as a {\it vertex operator\/} is exact (except, obviously, at the
  border of the wing, if the sequence is continued as \ $\ldots\to0$
  or \ $0\to\ldots$).  Thus, the mappings are exact everywhere except
  at the corner.  Next, the central mapping is given by the product of
  two screenings, and the cohomology can be worked out starting from
  the observation that $\Im \SB=\Im \SC =
  \dmV_{\frac{1-r}{p},p}(0,0,r-1)$, while $\Im \SB\circ \SC$ is the
  maximal submodule in~$\dmV_{\frac{1-r}{p},p}(0,0,r-1)$.  A more
  economical way to arrive at the same result is to map the butterfly
  resolution onto the two-sided resolution~\eqref{complex-top} as
  described in the next subsection.
\end{prf}

\subsection{Jamming the butterfly into the two-sided
  resolution\label{sec:jamming}} It is a reformulation of the fact
known since~\cite{[FMS]} that taking the kernel or the cokernel of the
fermionic screening gives rise to a $\beta\gamma$ system.  We apply
this to the butterfly resolution, and then the resulting $\beta\gamma$
system will be that of Sec.~\ref{sec:ghost} (with the butterfly
resolution becoming the ``linear'' two-sided
resolution~\eqref{complex-top}).  Indeed, since $Ce^{p\bar X}$ in
\eqref{two-fermionic} is a fermion, we define
\begin{equation}\label{eta-xi}
  \eta = C\,e^{p\bar X}\,,\qquad \xi = B\,e^{-p\bar X}
\end{equation}
and also introduce two scalars $\phi$ and $\varphi$ with signatures
$-1$ and $+1$, respectively:
\begin{equation}\label{phi}
  \d\phi=BC+\tfrac{1}{p}A - p\bar A\,,\qquad
  \d\varphi = BC + \tfrac{1}{p}A\,.
\end{equation}
The latter represents a fermionic ghost system $bc$ constructed as
in~\eqref{bc-ghosts}. Further, with $\eta$, $\xi$, and $\phi$
from~\eqref{eta-xi}--\eqref{phi}, we introduce a first-order bosonic
system in accordance with~\eqref{beta-gamma}.  Then the $\N2$
generators~\eqref{zeta-untwisted} can be rewritten in terms of these
$(b,c,\beta,\gamma)$ fields as
\begin{equation}
  \cQ=-\tfrac{1}{p}\beta\,c\,,\qquad\cG=(p-1)b\d\gamma
  -\gamma\d b\,,
\end{equation}
which differ from the respective generators in~\eqref{bcBG} only by
$\cQ\mapsto p\cQ$, $\cG\mapsto\frac{1}{p}\cG$, which does not change
the commutation relations (the rest of the algebra is {\it generated
  by\/} $\cQ$ and $\cG$).  Now the screening $\SC$ becomes simply
$\eta_0$, which ``reduces'' the $\eta\xi\phi$ space of states to the
$\beta\gamma$ space of states by eliminating the $\xi_0$ mode.

The other fermionic screening then becomes $\SB=\frac{1}{2\pi i}\oint
c\gamma^{p-1}$, which is~$\oQ_0$ from Sec.~\ref{sec:ghost}.

Now, the butterfly resolution can be mapped onto the two-sided
resolution~\eqref{complex-top}.  This goes by ``destroying'' the wings
such that only the $\SB$ mappings remain.  Consider first the right
wing.  We keep only the vertical border (see~\eqref{butterfly}):
setting $n=1$ in~\eqref{equality}, we are left with the mappings
\begin{equation}
  \dmU_{\frac{r+1}{p}-m-1,0,p;p-r}(p-r,-m,r-p-1)
  \xrightarrow{\SB}
  \dmU_{\frac{r+1}{p}-m,0,p;p-r}(p-r,-m+1,r-p-1)\,.
\end{equation}
In each module, we now take the quotient over the image of~$\SC$.  As
we have noted, this leaves us with the $bc\beta\gamma$ representation
space (i.e., the cohomology of \
$0\xleftarrow{\SC}\dots\xleftarrow{\SC}\dots$).  On the other hand,
the image of $\SC$ is the submodule in
$\dmU_{\frac{r+1}{p}-m-1,0,p;p-r}(p-r,-m,r-p-1)$ generated from
\begin{equation}\label{kernel}
  \ketBC{-mp}\tensor\ketAA{-m,r-p-1}
  \doteq\kettop{\tfrac{1-r}{p}+m,p;-p}\,.
\end{equation}
The quotient module is then generated from
\begin{equation}
  \ketBC{1-mp}\tensor\ketAA{-m,r-p-1}
  \doteq\kettop{\tfrac{1-r}{p}+m,p;-mp}
\end{equation}
(where $\doteq$ holds in the quotient, i.e. modulo the
vector~\eqref{kernel}).  This gives the mappings
\begin{multline}\label{right-half}
  \dots\xrightarrow{\SB}
  \pmV_{\frac{1-r}{p}+m+1,p;-(m+1)p}(1-(m+1)p,-m-1,r-p-1)
  \xrightarrow{\SB}\\
  \xrightarrow{\SB}
  \pmV_{\frac{1-r}{p}+m,p;-mp}(1-mp,-m,r-p-1)
  \xrightarrow{\SB}\dots
  \xrightarrow{\SB}
  \pmV_{\frac{1-r}{p}+1,p;0}(1-p,-1,r-p-1)\,,
\end{multline}
where $\pmV$ are the modules in which the singular vector
$\cG_{-mp}\,\ketBC{1-mp}\tensor\ketAA{-m,r-p-1}$ (see the top-left
extremal diagram in Fig.~\ref{fig:4}) has been factored over.

On the left wing, we take the modules on the vertical border ($n=0$
in~\eqref{left-B}--\eqref{left-C}) and keep only the kernel of~$\SC$
(again, the cohomology of \ 
$0\xrightarrow{\SC}\dots\xrightarrow{\SC}\dots$).  The kernel of~$\SC$
is generated from
\begin{equation}
  \ketBC{-r}\tensor\ketAA{m,r-1}
  \doteq\ket{\tfrac{r+1}{p}+m,0,p;-r}\,,\qquad m\geq0
\end{equation}
Denoting this module by
$\pmU_{m+\frac{r+1}{p},0,p;-r}(-r,m,r-1)$, we see that
mappings~\eqref{right-half} are continued as
\begin{equation}
  \pmU_{\frac{r+1}{p},0,p;-r}(-r,0,r-1)
  \xrightarrow{\SB}
  \pmU_{\frac{r+1}{p}+1,0,p;-r}(-r,1,r-1)
  \xrightarrow{\SB}
  \pmU_{\frac{r+1}{p}+2,0,p;-r}(-r,2,r-1)
  \xrightarrow{\SB}\dots
\end{equation}
We, thus, reproduce~\eqref{complex-top} transformed by the overall
spectral flow with~$\theta=1-r$.
\begin{Rem} Note that the {\it bosonic\/} screening of the $\N2$
  realisation~\eqref{zeta-untwisted}, which reads as
  \begin{equation}\label{bosonic-screening}
    S_{\mathrm{W}}=\tfrac{1}{2\pi i}\oint
    (p\bar A - BC)e^{-\frac{1}{p}X - \bar X},
  \end{equation}
  now takes the form~\eqref{Wak-screening}; it is {\it not\/} involved
  in the resolutions considered in this paper.
\end{Rem}

Under a further mapping to the Wakimoto representation, as we have
seen, the screening $\SC=\frac{1}{2\pi i}\oint\eta$, as before, serves
to correctly define the representation space of the first-order
bosonic system, now of the $\bbeta\ggamma$ system (thus, this
screening becomes redundant as soon as one deals with the $\bbeta$ and
$\ggamma$ fields as such, rather than with their bosonisation).  On
the other hand, as we have seen, the fermionic screening $\SB$ takes
the form~\eqref{QQ0-sl2} (note that the screenings finally take the
``standard'' vertex-operator form).

\section{Conclusions} We have constructed the two-sided ``linear''
resolution and the butterfly resolution of the unitary $\N2$
representations.  The resolutions show a ``strong dependence'' on the
free-field realisation chosen (moreover, the ``asymmetric''
realisation~\cite{[GS2],[BLNW]} of the $\N2$ algebra shows yet
different resolution structures~\cite{[ST-un]}).  In fact, it is the
quantum group encoded in the structure of the screenings that
determines the resolution.  In the $bc\beta\gamma$ bosonisation of the
$\N2$ algebra, we have seen (Theorem~\ref{thm:Q0}) to what extent the
folklore statement that ``the cohomology is generated from chiral
primary fields'' is true: the cohomology does contain a chiral primary
state (the topological \hw{} states as we call them here) for each $r$
from $1\leq r\leq p-1$, however these, on the one hand, can also be
{\it twisted\/} by the spectral flow, and on the other hand, have to
correspond to a very particular irreducible subquotient
in~\eqref{pattern-first}
%the occupy a very special position in diagram~\eqref{perverted-emb}
(where infinitely many other twisted topological \hw{} states are {\it
  not\/} in the cohomology).

As regards the butterfly resolution, its shape can be heuristically
interpreted as a ``Felder-type'' effect occurring in the
``$3,5,7,.\,.\,.$''-resolution of irreducible $\N2$
representations~\cite{[FSST]}.  The latter is constructed in terms of
twisted {\it massive\/} Verma modules and goes like
\begin{equation}\label{massive-resolution}
  0\leftarrow\bullet\leftarrow\bullet
    \leftarrow
  \begin{array}{l}
    \bullet\\
    \bullet\\
    \bullet
  \end{array}
    \leftarrow
  \begin{array}{l}
    \bullet\\
    \bullet\\
    \bullet\\
    \bullet\\
    \bullet
  \end{array}
  \leftarrow\dots
\end{equation}
Note that folding (somewhat asymmetrically) the butterfly's wings
reproduces the pattern of~\eqref{massive-resolution}.  Thus, the moral
is that the $bc\beta\gamma$ realisation of $\N2$ modules turns some of
the mappings in~\eqref{massive-resolution} ``inside out,'' thus
resulting in~\eqref{butterfly-0}.  The Felder resolution can be viewed
similarly, with the ``braid'' of type~\eqref{standard-emb} becoming an
infinite line with the cohomology in the center.  This has been given
a precise meaning in~\cite{[FFr]}; an interesting question, therefore,
is about the construction of the ``intermediate'' modules that
interpolate between the $3,5,7,.\,.\,.$- and the butterfly
resolutions.

As follows from the remarks made in the text, the butterfly resolution
of the unitary $\tSL2$ representations involves twisted (spectral-flow
transformed) Wakimoto modules; with the $\bbeta\ggamma$ pictures being
different for different modules, this resolution can be viewed as
pertaining to the three-boson realisation obtained by additionally
bosonising the $\bbeta\ggamma$ fields in the Wakimoto representation.

{}From the LG perspective, relation~\eqref{G(z)-relation} which
characterises the unitary $\N2$ representations, demonstrates a formal
similarity with the unperturbed $A_{p-1}$ LG equations of motion
(suppressing the kinetic term) $X^{p-1}=0$,
Eq.~\eqref{gamma-constraint}; to continue with the parallel, moreover,
recall that the chiral ring in the LG description is generated by $1$,
$X$, \dots, $X^{p-2}$; it turns out that the unitary $\N2$
representations can be spanned by acting {\it solely with the modes
$\cG_{n}$, $n\in\oZ$}, subject to the constraints following
from~\eqref{G(z)-relation}, with no other $\N2$ generators
involved~\cite{[FST-semi]}.  Such a characterization of irreducible
representations by a fermionic counterpart of the LG equations of
motion would be extremely interesting to generalize to the case
involving more than one ghost system of each sort.

\bigskip

\noindent{\it Acknowledgements}.  We are grateful to I.~Tipunin
for very useful discussions.  AMS is also grateful to V.~Schomerus and
K.~Sfetsos for discussions on some related subjects, and to K.~Sfetsos
for pointing out the paper~\cite{[DLP]}, in which several aspects of a
later construction of~\cite{[FST]} were anticipated.  This work was
supported in part by the RFBR Grant 98-01-01155.
%%96-02-16117. %%Fainberg

\appendix
\section{$\N2$ Verma modules\label{sec:top-Verma}} We first introduce
the class of $\N2$ Verma modules that we call {\it
  topological\/}\footnote{chiral, in a different set of conventions,
  see, e.g.,~\cite{[LVW]}.} Verma modules
following~\cite{[FST],[ST4]}.  For a fixed $\theta\in\oZ$, we define
the {\it twisted topological \hw{} vector\/} $\kettop{h,t;\theta}$ to
satisfy the annihilation conditions (which are referred to as the
twisted topological \hw{} conditions)
\begin{equation}
  \cQ_{-\theta+m}\kettop{h,t;\theta}=
  \cG_{\theta+m}\kettop{h,t;\theta}=
  \cL_{m+1}\kettop{h,t;\theta}=
  \cH_{m+1}\kettop{h,t;\theta}=0\,,\quad m\in\oN_0\,,
  \label{annihiltop}
\end{equation}
with the following eigenvalues of the Cartan generators (where the
second equation follows from the annihilation conditions):
\begin{align}
    (\cH_0+\tfrac{\ctop}{3}\theta)\,\kettop{h,t;\theta}&=
    h\,\kettop{h,t;\theta}\,,\label{H0-top}\\
    (\cL_0+\theta\cH_0+\tfrac{\ctop}{6}(\theta^2+\theta))
    \,\kettop{h,t;\theta}&=0\,.
    \label{L0-top}
\end{align}
The parameter $t$ fixes (the eigenvalue of) the central charge as
$\ctop=3(1-\frac{2}{t})$.
\begin{Dfn}\label{dfn:top}
  The {\it twisted topological Verma module\/} $\mV_{h,t;\theta}$ is
  the module freely generated from the topological \hw{} vector
  $\kettop{h,t;\theta}$ by $\cQ_{\leq-1-\theta}$,
  $\cG_{\leq-1+\theta}$, $\cL_{\leq-1}$, and $\cH_{\leq-1}$.
\end{Dfn}

We write $\kettop{h,t}\equiv \kettop{h,t;0}$ in the `untwisted' case
of $\theta=0$ and also denote by $\mV_{h,t}\equiv\mV_{h,t;0}$ the
untwisted module.

Submodules in a topological Verma modules are twisted topological
Verma modules (or a sum of two such modules).  A singular vector
exists in $\smV_{h,t;\theta}$ if and only if $h=\hplus(r,s,t)$ or
$h=\hminus(r,s,t)$, where
\begin{equation}
  \begin{split}
    \hplus(r,s,t)&=\tfrac{1-r}{t}+s-1\,,\\
    \hminus(r,s,t)&=\tfrac{1+r}{t}-s\,,
  \end{split}
  \quad r,s\in\oN\,.
  \label{hplusminus}
\end{equation}
We denote these singular vectors $\ket{E(r,s,t)}^{\pm,\theta}$,
respectively (omitting the twist $\theta$ when it is equal to zero).
The submodule of $\smV_{h,t;\theta}$ generated from
$\ket{E(r,s,t)}^{\pm,\theta}$ is the twisted topological Verma module
$\smV_{h\pm r\frac{2}{t},t;\theta\mp r}$.  When $s=1$, the topological
singular vectors take a particularly simple form,
\begin{align}
  \ket{E(r,1,t)}^{+,\theta}&=\cG_{\theta-r}\dots\cG_{\theta-1}\,
  \kettop{\hplus(r,1,t),t;\theta}\,,\label{Eplus}\\
  \ket{E(r,1,t)}^{-,\theta}&=\cQ_{-\theta-r}\dots\cQ_{-\theta-1}\,
  \kettop{\hminus(r,1,t),t;\theta}\,,\label{Eminus}
\end{align}
while for $s\geq2$ singular vectors in topological Verma modules are
given by the construction of~\cite{[ST3],[ST4]}.  This involves the
continued operators $q(a,b)$ and $g(a,b)$ that become the products
$\cQ_a\cQ_{a+1}\dots\cQ_b$ and $\cG_a\cG_{a+1}\dots\cG_b$,
respectively, whenever $b-a+1\in\oN$. In terms of these, the singular
vectors read as
\begin{align}
  \ket{E(r,s,t)}^+&=
  g(-r,(s-1)t-1)\,\cE^{-,(s-1)t-r}(r,s-1,t)\,
  g((s-1)t-r,-1)\,\kettop{\hplus(r,s,t),t}\,,\label{Tplus}\\
  \ket{E(r,s,t)}^-&=
  q(-r, (s-1) t - 1)\,\cE^{+,r-(s-1)t}(r,s-1,t)\,
  q((s-1) t - r, -1)\,\kettop{\hminus(r,s,t),t}\,,\label{Tminus}
\end{align}
where $\cE^{\pm}$ are the corresponding singular vector {\it
  operators\/} and $\cE^{\pm,\theta}$ is their spectral flow transform
by~$\theta$.  There exists a set of algebraic rules~\cite{[ST4]} that
allow one to evaluate these expressions as monomials in the~$\N2$
generators $\cQ_n$, $\cG_n$, $\cH_n$, and $\cL_n$, $n\in\oZ$, once the
singular vectors with $s\mapsto s-1$ are evaluated in this form; with
the $s=1$ vectors given by~\eqref{Eplus} and \eqref{Eminus}, this
gives a recursive procedure to evaluate all singular vectors in
(twisted) topological Verma modules~\cite{[ST3]}.

For $t=p\in\oN+2$ and $1\leq r\leq p-1$, the topological Verma
module\footnote{Where we start with the module twisted by $r-1$
  because we need this in Sec.~\ref{sec:ghost}; the overall twist can
  be applied to~\eqref{standard-emb} straightforwardly,
  see~\cite{[FSST]}.}  $\smV_{\frac{1-r}{p}, p;r-1}$ has the following
embedding diagram~\cite{[SSi],[FSST]} consisting of twisted
topological Verma modules, for which we indicate the $h$ and $\theta$
values as \ $(h;\theta)$:
\begin{equation}\label{standard-emb}
  \unitlength=1pt
  \begin{picture}(400,100)
    \put(0,20){
      \put(40,0){
        \put(0,30){$\bullet$}
        \put(6,29){\vector(1,-1){23}}
        \put(6,36){\vector(1,1){23}}
        \multiput(30,0)(60,0){2}{
          \put(0,0){$\bullet$}
          \put(0,60){$\bullet$}
          \put(7,7){\vector(1,1){50}}
          \put(7,57){\vector(1,-1){50}}
          \put(7,3){\vector(1,0){50}}
          \put(7,63){\vector(1,0){50}}
          }
        \multiput(160,3)(170,0){2}{\Large\dots}
        \multiput(160,61)(170,0){2}{\Large\dots}
        }
      \put(0,20){${}^{(\frac{1-r}{p};r-1)}$}
%%lower line:
      \put(50,-17){${}^{(\frac{r+1}{p};-1)}$}
      \put(110,-17){${}^{(\frac{1-r}{p}+2;r-1-p)}$}
      \put(220,-17){$\atop{(\frac{r+1}{p}+2(m-1);}{\quad~~(1-m)p-1)}$}
      \put(290,-17){${}^{(\frac{1-r}{p}+2m;r-1-mp)}$}
%%upper line:
      \put(50,64){${}^{(\frac{r+1}{p}-2;p-1)}$}
      \put(110,64){${}^{(\frac{1-r}{p}-2;p+r-1)}$}
      \put(220,64){${}^{(\frac{r+1}{p}-2m;mp-1)}$}
      \put(290,64){${}^{(\frac{1-r}{p}-2m;mp+r-1)}$}
      \multiput(240,0)(60,0){2}{
        \put(0,0){$\bullet$}
        \put(0,60){$\bullet$}
        \put(7,57){\vector(1,-1){50}}
        \put(7,7){\vector(1,1){50}}
        \put(7,3){\vector(1,0){50}}
        \put(7,63){\vector(1,0){50}}
        }
      }
  \end{picture}
\end{equation}
The arrows are drawn in the direction of {\it sub\/}modules.  For
Verma modules, such a digram represents the hierarchy of singular
vectors and hence the ``adjacency'' of the irreducible subquotients.
For Wakimoto-like modules, it is more natural to view similar diagrams
as showing the adjacency of subquotients: every dot is an {\it
  irreducible\/} subquotient, and an arrow
${\stackrel{A}{\bullet}}\to{\stackrel{B}{\bullet}}$ means that there
exists a module in which $B$ is a submodule and $A$ is the quotient
(this is often expressed by saying that $A$ is glued to~$B$); thus,
the arrows represent all the subquotients consisting of two
irreducible ones.
%In this language, for example, the Wakimoto modules over $\tSL2$
%differ from the Verma modules by their ``adjacency relations,'' i.e.,
%by how the composition series factors (which are the same in both
%cases) are glued together.

\bigskip

A different class of Verma-like $\N2$ modules are defined as
follows~\cite{[ST4]}.
\begin{Dfn}
  A twisted massive Verma module $\mU_{h,\ell,t;\theta}$ is freely
  generated from a twisted {\it massive \hw{} vector\/}
  $\ket{h,\ell,t;\theta}$ by the generators
  \begin{equation}
    \cL_{-m}\,,~m\in\oN\,,\qquad \cH_{-m}\,,~m\in\oN\,,\qquad
    \cQ_{-\theta-m}\,,~m\in\oN_0\,,\qquad \cG_{\theta-m}\,,~m\in\oN\,.
    \label{verma}
  \end{equation}
  The twisted massive \hw{} vector $\ket{h,\ell,t;\theta}$ satisfies
  the following conditions:
  \begin{gather}\label{masshw}
    \cQ_{m+1-\theta}\,\ket{h,\ell,t;\theta}=
    \cG_{m+\theta}\, \ket{h,\ell,t;\theta}=
    \cL_{m+1}\,\ket{h,\ell,t;\theta}=
    \cH_{m+1}\,\ket{h,\ell,t;\theta}=0\,,~m\in\oN_0\,,\\
    \begin{aligned}
      (\cH_0 + \tfrac{\ctop}{3}\theta)\,\ket{h,\ell,t;\theta}&=
      h\,\ket{h,\ell,t;\theta}\,,\\
      (\cL_0 + \theta\cH_0 + \tfrac{\ctop}{6}(\theta^2 +
      \theta))\,\ket{h,\ell,t;\theta} &=
      \ell\,\ket{h,\ell,t;\theta}\,.
    \end{aligned}\label{masshw-2}
  \end{gather}
\end{Dfn}
\noindent
We also write $\ket{h,\ell,t}=\ket{h,\ell,t;0}$ and
$\mU_{h,\ell,t}=\mU_{h,\ell,t;0}$.

A {\it charged singular vector\/} occurs in $\mU_{h,\ell,t;\theta}$
whenever~\cite{[BFK]}
\begin{equation}\label{ell-ch}
  \ell=\ellch(n,h,t)\equiv -n(h-\tfrac{n+1}{t})\,,\quad
  n\in\oZ\,,
\end{equation}
and reads as~\cite{[ST3],[ST4]}
\begin{equation}
  \ket{E(n,h,t)}_{\mathrm{ch}}^{\theta}=\left\{
    \begin{aligned}
      \cQ_{-\theta-n}\,\ldots\,\cQ_{-\theta}\,
      \ket{h,\ellch(n,h,t),t;\theta}\,,
      &\quad&n\geq0\,,\\
      \cG_{\theta+n}\,\ldots\,\cG_{\theta-1}\,
      \ket{h,\ellch(n,h,t),t;\theta}\,,&&
      n\leq-1\,.
    \end{aligned}\right.
  \label{ECh}
\end{equation}
This state on the extremal diagram satisfies the {\it twisted\/}
topological \hw{} conditions with the twist~$\theta+n$, the submodule
generated from $\ket{E(n,h,t)}_{\mathrm{ch}}^\theta$ being the twisted
topological Verma module $\mV_{h-\frac{2n}{t}-1,t;n+\theta}$ if
$n\geq0$ and $\mV_{h-\frac{2n}{t},t;n+\theta}$ if $n\leq-1$.  It is
useful to represent the charged singular vectors, for $n\geq0$ and
$n\leq-1$ respectively, as follows:
\begin{equation}\label{charged-picture}
  \begin{picture}(400,80)
    \put(0,-30){
      \qbezier(-10,42)(60,102)(160,98)
      \put(134,102){\scriptsize$\cQ_{-n}$}
      {\thicklines\put(136,98){\vector(1,0){14}}}
      \qbezier(150,98)(130,95)(110,80)
      }
    \put(200,-30){
      \qbezier(-10,42)(60,102)(160,98)
      \put(-4,58){\scriptsize$\cG_{n}$}
      {\thicklines\put(12,58){\vector(-3,-2){12}}}
      \qbezier(0,50)(22,56)(52,52)
      }
  \end{picture}
\end{equation}
Here, the arrows cannot be inverted by acting with the opposite mode
of the other fermion from the $\N2$ algebra -- which precisely means
that the state obtained satisfies twisted {\it topological\/} \hw{}
conditions and generates a {\it sub\/}module.  In the body of the
paper, we often deal with the charged singular vectors with $n=0$,
which exist in~$\mU_{h,0,p;\theta}$, and those with $n=-1$
in~$\mU_{h,h,p;\theta}$.  Note finally that whenever a charged
singular vector generates a maximal submodule (i.e., is not contained
in a submodule generated from another charged singular vector), the
quotient of the massive Verma module over the corresponding submodule
is a twisted topological Verma module.

\end{document}